\documentclass[final]{siamltex}

\usepackage{amsmath}
\usepackage{amssymb}
\usepackage{graphicx}
\usepackage{array}

\newtheorem{myguide}{Guideline}

\begin{document}

\title{High-Performance Solvers for Dense Hermitian
  Eigenproblems\thanks{Financial support  
from the Deutsche Forschungsgemeinschaft (German Research Association)
through grant GSC 111 is gratefully acknowledged.
}} 


\author{
  M.~Petschow\thanks{Aachen Institute for advanced study in Computational 
Engineering Science, RWTH Aachen University (\{petschow,peise,pauldj\}@aices.rwth-aachen.de).}
  \and E.~Peise\footnotemark[2]
  \and P.~Bientinesi\footnotemark[2]
}

\maketitle  

\begin{abstract}

  We introduce a new collection of solvers -- subsequently called EleMRRR --
  for large-scale dense Hermitian eigenproblems.
  EleMRRR solves various types of problems:
  generalized, standard, and tridiagonal
  eigenproblems.
  Among
  these, the last is of particular importance as it is a solver on
  its own right, as well as the computational kernel for the first 
  two; we present a fast and scalable 
  tridiagonal solver based on the algorithm of
  {\it Multiple Relatively Robust Representations} -- referred to as PMRRR. 
  Like the other EleMRRR solvers, PMRRR is part of the freely available
  Elemental library, and is designed to fully support both
  message-passing (MPI) and multithreading parallelism (SMP). As a
  result, the solvers are equally effective in message-passing environments
  with and without SMP parallelism. 
  We conducted a thorough performance study of EleMRRR and  ScaLAPACK's
  solvers on two supercomputers. 
  Such a study, performed with up to $8{,}192$ cores,  
  provides precise guidelines 
  to assemble the fastest solver within the ScaLAPACK framework;
  it also indicates that EleMRRR outperforms even the fastest solvers
  built from ScaLAPACK's components.

\end{abstract}



\pagestyle{myheadings}
\thispagestyle{plain}
\markboth{M.~PETSCHOW, E.~PEISE, AND P.~BIENTINESI}{LARGE-SCALE DENSE EIGENSOLVERS}

\section{Introduction}
\label{Introduction}

In this section we briefly state the considered eigenproblems, give a
short description of our high-level approach, and list the
main contributions of this paper.

\subsection{The Problem}
\label{theproblem}
A {\em generalized
  Hermitian\footnote{That is $A = A^*$ and $B = B^*$. 
    If both Hermitian matrices $A$ and $B$ are real valued, 
    then the following discussion also holds with the words
  'Hermitian' and 'unitary' respectively replaced by 'symmetric' and
  'orthogonal'.} eigenproblem} (GHEP) is identified by the equation 
\begin{equation}
  A x = \lambda B x \,,   \label{eq:GenEig}
\end{equation}
where $A, B \in \mathbb{C}^{n\times n}$ are known matrices; the sought after
scalars $\lambda$ and associated 
vectors $x \in \mathbb{C}^{n}$, $x \neq 0$, are called  
{\em eigenvalues} and {\em eigenvectors}, respectively. We say that
$(\lambda, x)$ is an {\em eigenpair} of the {\em pencil} $(A,B)$. 
In the following, we
 make the 
additional assumption that either $A$ or $B$ is positive or negative
definite, which implies that all the eigenvalues are real. Without loss of
generality, we assume that $B$ is positive 
definite. Thus, when referring to the GHEP of Eq.~\eqref{eq:GenEig}, the
restriction to Hermitian-definite pencils $(A,B)$ is subsequently implied. 
If $B$ is the identity matrix $I$, Eq.~\eqref{eq:GenEig} reduces to 
the {\em standard Hermitian eigenproblem} (HEP) $A x =
\lambda x$; 
if $A$ is also real and tridiagonal, the problem is referred to as a 
{\em symmetric tridiagonal eigenproblem} (STEP). 

Different numerical methods were devised to solve instances of
the GHEP, HEP, and STEP in accordance to the amount of
eigenpairs requested 
and additional properties of the involved matrices (e.g., sparsity). In
this paper, we concentrate on the efficient solution of {\em dense}
generalized eigenproblems on modern distributed-memory
platforms composed of shared-memory multiprocessor nodes; as a by-product,
we also obtain results for standard and tridiagonal problems.



\subsection{Three Nested Eigensolvers}
\label{nestedsolvers}
Common methods to the dense generalized eigenproblems make
use of the following fact: given non-singular matrices $G$ and $F$,
the eigenvalues of the pencil $(A,B)$ are invariant under the equivalence
transformation $(GAF, GBF)$; furthermore, $x$ is an eigenvector of
$(A,B)$ if and only if $F^{-1}x$ is an eigenvector of $(GAF,
GBF)$~\cite{Parlett:1998:SEP:280490}.

The most versatile tool for the generalized eigenproblem, the QZ algorithm~\cite{Moler73}, uses a sequence of
unitary 
equivalence transformations to reduce the original pencil to generalized
(real) Schur form. By
design, the  
QZ algorithm is numerically backward stable and imposes no restrictions on the
input matrices; unfortunately, the algorithm does not respect the symmetry
of the Hermitian pencil $(A,B)$ and is computationally rather
costly.

To preserve the symmetry of the problem while reducing the pencil
$(A,B)$ to simpler form, methods are limited to sequences of
congruence transformations -- that is, using $G = F^*$, where $G$ and $F$ are no
longer required to be unitary.   
The traditional approach for computing all or a
    significant fraction of the eigenpairs 
of $(A,B)$ relies on a
reduction-backtransformation procedure, corresponding to three nested
eigensolvers: generalized, standard, and tridiagonal.
Since $B$ is positive definite, the Cholesky factor can be used to transform
the GHEP to a HEP~\cite{Martin68}, 
which in turn is reduced to a STEP;
once the eigenvectors of the STEP are computed, 
they are mapped back to those of the original problem 
via two successive backtransformations. 
Overall, the process for solving a generalized eigenproblem -- also known as
the Cholesky-Wilkinson method -- consists
of six stages: 
\begin{enumerate}
\item {\it Cholesky factorization:}
$B$ is factored into $L L^*$, where $L$ is lower triangular.

\item 
  {\it Reduction to standard form:} 
From $B = L L^*$, the original pencil $(A,B)$ is transformed to $(L^{-1} A
L^{-*},I)$ and Eq.~\eqref{eq:GenEig} takes the form of a standard Hermitian
eigenproblem $ M y = \lambda y$. 
The eigenvectors $x$ of the GHEP and the
eigenvectors of the HEP are related by $y = L^* x$.
\item 
  {\it Reduction to tridiagonal form:} 
A unitary matrix $Q$ is computed such that
$T = Q^* M Q$ is real symmetric tridiagonal. The pencil $(M, I)$ is
transformed to $(Q^* M Q, I)$, corresponding to the tridiagonal eigenproblem $T
z = \lambda z$, where $z =
Q^* y$.
\item {\it Solution of the tridiagonal eigenproblem:}
A set of eigenpairs $(\lambda, z)$ is computed such that
$T z = \lambda z$; the existence of $n$ such eigenpairs $(\lambda, z)$ for
which the set of eigenvectors forms an orthonormal basis for $\mathbb{C}^n$
is ensured by the Spectral Theorem~\cite{Parlett:1998:SEP:280490}.

\item {\it First backtransformation:} In accordance to Stage 3, the 
  eigenvectors of the standard eigenproblem are obtained by computing $y = Q
  z$. 

\item {\it Second backtransformation:}
In accordance to Stage 2, the
eigenvectors of the original pencil are obtained by computing $x = L^{-*} y$.
\end{enumerate}
The above discussion shows that when the $k$ computed
eigenvalues are the entries of a diagonal matrix $\Lambda \in
\mathbb{R}^{k \times k}$, and the associated 
eigenvectors are the columns of a matrix $X \in \mathbb{C}^{n \times k}$, then
 $X^* A X = \Lambda$, $X^* B X = I$, and $A X = B X \Lambda$. 

With slight modifications of Stages 2 and 6, the same six-stage procedure
also applies to eigenproblems in the form $A B x=\lambda x$ and
$ B A x=\lambda x$. In the first case, the reduction and the
backtransformation become $ M = L^* A L $ and $ x = L^{-*} y $,
respectively; in the second case, they become $ M = L^* A L $
and $ x = L y $.

The six-stage procedure should only be used if $B$ is sufficiently
well-conditioned with respect to inversion. For a detailed discussion of the
properties of the GHEP, especially regarding perturbation theory and the
problems arising for ill-conditioned $B$, we refer to the standard literature,
including~\cite{Bai:2000:TSA:357352,Golub1996,Parlett:1998:SEP:280490,stewart-sun:1990}.

\subsection{The Impact and Contributions of this Paper}
The solution of the
aforementioned types of large-scale eigenproblems is  
integral to a number of scientific
disciplines, particularly vibration analysis 
\cite{Bennighof04} and 
quantum chemistry~\cite{Hafner20076,Persson2007280,doi:10.1002/jcc.20549,doi:10.1021/ct900539m}. 
An example of an application where the solution of the eigenproblem is
the most time consuming operation is Density Functional
Theory~\cite{kent2008,doi:10.1021/ct900539m,tomic2006}. There, as 
part of a simulation, one has to solve a set of equations in a
self-consistent fashion; this is accomplished by an iterative process 
in which each iteration involves the solution of
dozens or even hundreds of generalized eigenproblems.
In many cases, the application requires 5-25\% of the eigenpairs
  associated with the smallest eigenvalues, and 
the size of such problems is usually in the tens of thousands of degrees of
freedom. 
In other applications, in which the simulations do not follow an iterative process, 
it is instead common to encounter only one single eigenproblem --
potentially of very large size 
(50k--100k or more)~\cite{MaeDiss04,BioApp}. 
In both scenarios, the problem size is not limited by
the physics of the problem, but only by memory requirements and
time-to-solution.

When the execution time and/or the memory requirement of a simulation
become limiting factors, scientists place their hopes on massively
parallel supercomputers.  
With respect to execution time, the use of
more processors would ideally result in faster solutions.  When memory
is the limiting factor, additional resources from large
distributed-memory environments should enable the solution of larger
problems.
We study the performance of eigensolvers for both
situations: increasing the number of processors while keeping the
problem size constant ({\em strong scaling}), and increasing the
number of processors while keeping the memory per node constant ({\em
  weak scaling}). 


In this paper we make the following contributions:
\begin{itemize}
\item
Given the nested nature of the generalized, standard and tridiagonal
eigensolvers, the last is both a
solver in its own right and the computational kernel for the HEP and
the GHEP. We present a novel tridiagonal solver, PMRRR, based on the algorithm
of Multiple Relatively Robust Representations
(MRRR)~\cite{DhillonDiss97,Dhillon04multiplerepresentations}, which 
merges the distributed and multithreaded approaches 
first introduced in \cite{Bientinesi:2005:PED:1081198.1081222}
and~\cite{mr3smp}.\footnote{PMRRR should not be confused with the
  distributed-memory solver introduced
  in~\cite{Bientinesi:2005:PED:1081198.1081222}.} PMRRR is well suited for
both single node 
and large scale massively parallel computations. Experimental results
indicate that PMRRR is currently the fastest tridiagonal solver available,
outperforming all the solvers included in LAPACK~\cite{laug}, ScaLAPACK
\cite{Dongarra:1997:SUG:265932} and Intel's Math Kernel Library (MKL). 

\item
We introduce EleMRRR (from Elemental and PMRRR), a set of
distributed-memory 
eigensolvers for generalized and standard Hermitian eigenproblems. EleMRRR
provides full support for hybrid message-passing and
multithreading parallelism. If multithreading is not desired, EleMRRR can be
used in purely message-passing mode. 
%

\item
The five stages of reduction and backtransformation in EleMRRR are based on
Elemental, a library for the development of distributed-memory dense linear
algebra routines~\cite{ElemTOMS}.  
Elemental embraces a two-dimensional cyclic 
element-wise matrix distribution, and attains performance comparable or even
superior to the well established ScaLAPACK, PeIGS and PLAPACK parallel libraries
\cite{Dongarra:1997:SUG:265932,dhillonfannparlett1,PLAPACKbook}. For the reduction to standard
form, an algorithmic variant is 
used which delivers high performance and scalability~\cite{Elemental2:TOMS}. 

\item
A thorough performance study on two high-end computing platforms is
provided. This study accomplishes two objectives.  On the one hand it
contributes guidelines on how to build -- within the ScaLAPACK
framework -- an eigensolver faster than the existing ones.  This is of
particular interest to computational scientists and
engineers as each 
of the commonly used\footnote{See for example \cite{doi:10.1021/ct900539m,dai2008,kent2008,tomic2006}.} routines ({\tt PZHEGVX}, {\tt PDSYGVX}, {\tt
  PZHEEVD}, {\tt PDSYEVD}) present
performance penalties 
that can be avoided by calling a different
sequence of subroutines and choosing suitable settings.  
On the other hand, the study indicates that
EleMRRR is scalable -- both strongly and weakly -- to a large number of
processors, and outperforms the standard ScaLAPACK solvers. 
\end{itemize}

The paper is organized as follows: In Section~\ref{relatedwork} we
discuss related work and give experimental evidence that some widely
used routines fail to deliver the desired performance. In
Section~\ref{elemental}, we
concentrate on EleMRRR, with emphasis on the
tridiagonal stage -- PMRRR. We present a thorough performance study on two
state-of-the-art high-performance computer systems in Section
\ref{experiments}. We show that 
ScaLAPACK contains a set of fast routines that can be combined to avoid the
aforementioned performance problems, and we compare the resulting routines
to our solver EleMRRR. We summarize our findings in Section~\ref{conclusions}.

\section{A Study of Existing Solvers}
\label{relatedwork}



In this section we give a brief overview of existing methods and study
well-known issues of widely used routines available in the current
version\footnote{At the time of writing, ScaLAPACK's
  latest version was 1.8. The version, 2.0, presents no
  significant changes in the tested routines.} of the ScaLAPACK library. We
discuss the generalized, standard and symmetric tridiagonal eigenproblems in
succession.

\subsection{The Generalized Eigenproblem} 
\label{geneig}


In some cases, even if $A$ and $B$ do not satisfy the
assumptions of Section~\ref{theproblem}, the problem can still be transformed
into one that exhibits the desired
properties~\cite{Bai:2000:TSA:357352}.
In general, if $A$ and $B$ are dense but non-Hermitian or if $B$ has poor
conditioning with respect to inversion\footnote{If $B$ is (nearly)
  semi-definite, instead of the QZ algorithm, one can use 
  a variant of the reduction to standard form introduced by \cite{FH:SINum:72}. However, in
  contrast to the QZ algorithm, this variant is not included in any of the most 
  widely used libraries.}, instead of
the aforementioned 
six-stage approach, the QZ algorithm 
can be used.
A parallel distributed-memory
implementation is discussed in~\cite{Adlerborn2006}.


The ScaLAPACK library 
contains routines for the three classes of eigenproblems 
we consider in this paper. 
A complete list of routine names for both the solvers and the individual 
stages is given in Table~\ref{scalapackroutines}.
In particular, {\tt
PZHEGVX} and {\tt 
  PDSYGVX} are ScaLAPACK's double precision drivers for complex
Hermitian and real symmetric generalized eigenproblems, respectively. 

\begin{table}[htb]
\begin{center}
\footnotesize
\begin{tabular}{ c  l  l  l  } \hline\noalign{\smallskip}
  Stage & Complex        & Real  & Description \\ \hline\hline\noalign{\smallskip}
  1---6  & {\tt PZHEGVX}  & {\tt PDSYGVX} & GHEP --- Bisection and Inverse Iteration \\
  3--5  & {\tt PZHEEVX}  & {\tt PDSYEVX} & HEP --- Bisection and Inverse Iteration \\
  3--5  & {\tt PZHEEV}   & {\tt PDSYEV}  & HEP --- QR algorithm
  \\
  3--5  & {\tt PZHEEVD}  & {\tt PDSYEVD} & HEP ---
  Divide and Conquer algorithm \\
  3--5  & {\tt PZHEEVR}  & {\tt PDSYEVR} & HEP --- MRRR
  algorithm \\
  4     & - & {\tt PDSTEDC} & STEP ---
  Divide and Conquer algorithm \\
  4     & - & {\tt PDSTEGR} & STEP --- MRRR
  algorithm \\[1mm]
  \hline\noalign{\smallskip} 
  1     & {\tt PZPOTRF}  & {\tt PDPOTRF} & Cholesky factorization \\
  2     & {\tt PZHEGST}  & {\tt PDSYGST} & Reduction to HEP \\
  2     & {\tt PZHENGST} & {\tt PDSYNGST}& Reduction to HEP (square grid of processes) \\
  3     & {\tt PZHETRD}  & {\tt PDSYTRD} & Reduction to STEP \\
  3     & {\tt PZHENTRD} & {\tt PDSYNTRD}& Reduction to STEP (uses {\tt
    PxxxTTRD} for square grid of
  processes) \\
  4     & - & {\tt PDSTEBZ} & Eigenvalues of a tridiagonal matrix using Bisection\\
  4     & {\tt PZSTEIN} & {\tt PDSTEIN} & Eigenvectors of a tridiagonal matrix using Inverse Iteration\\
  5     & {\tt PZUNMTR}  & {\tt PDORMTR} & Backtransformation to HEP \\
  6     & {\tt PZTRSM}   & {\tt PDTRSM}  & Backtransformation to GHEP (Type \#1 or \#2) \\
  6     & {\tt PZTRMM}   & {\tt PDTRMM}  & Backtransformation to GHEP (Type \#3) \\[1mm]
  \hline 
\end{tabular}
\end{center}
\caption{List of relevant ScaLAPACK routine names. 
}
\label{scalapackroutines}
\end{table}

\begin{figure}[t]
   \centering
\includegraphics[scale=.38]{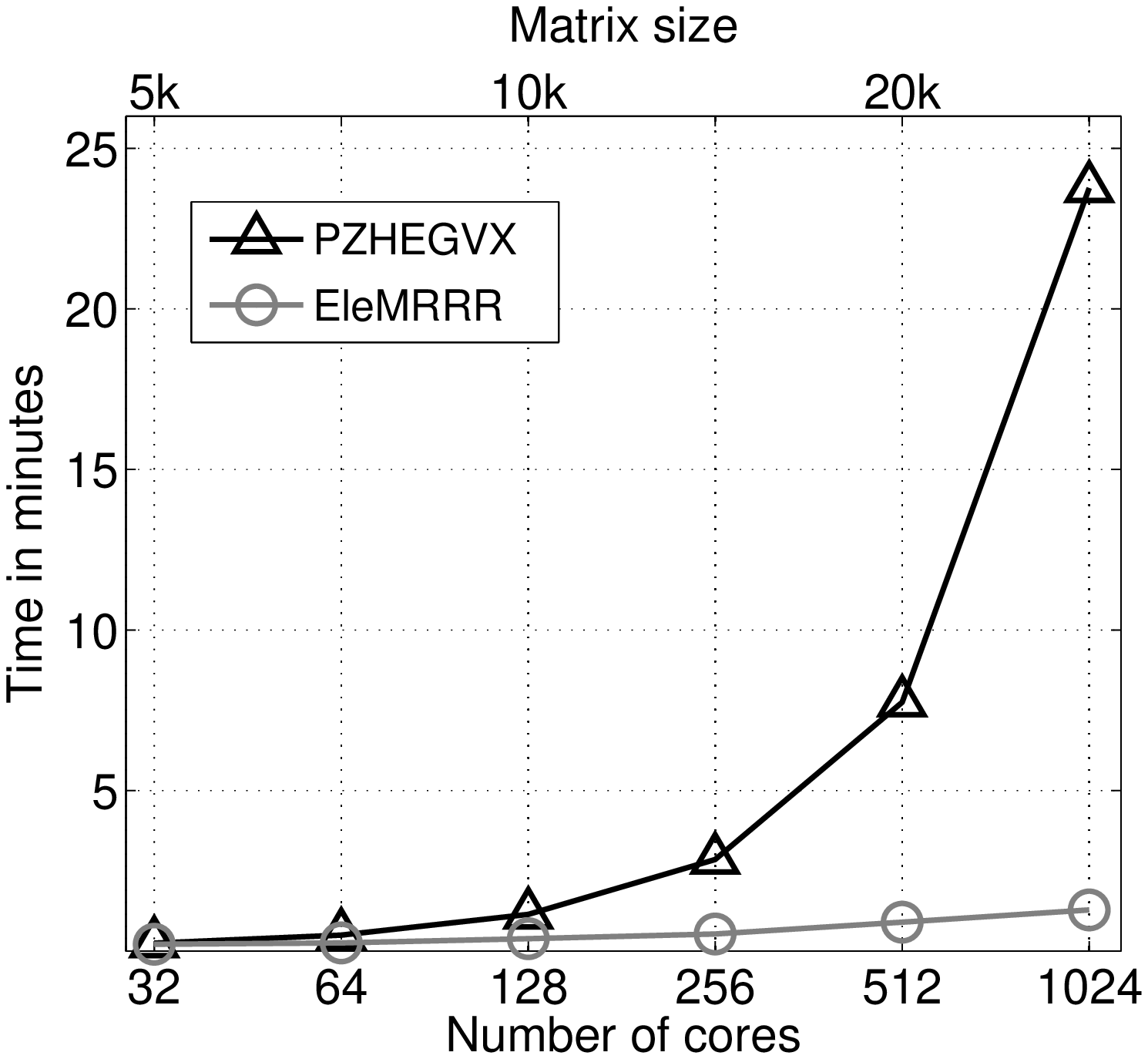}
\includegraphics[scale=.38]{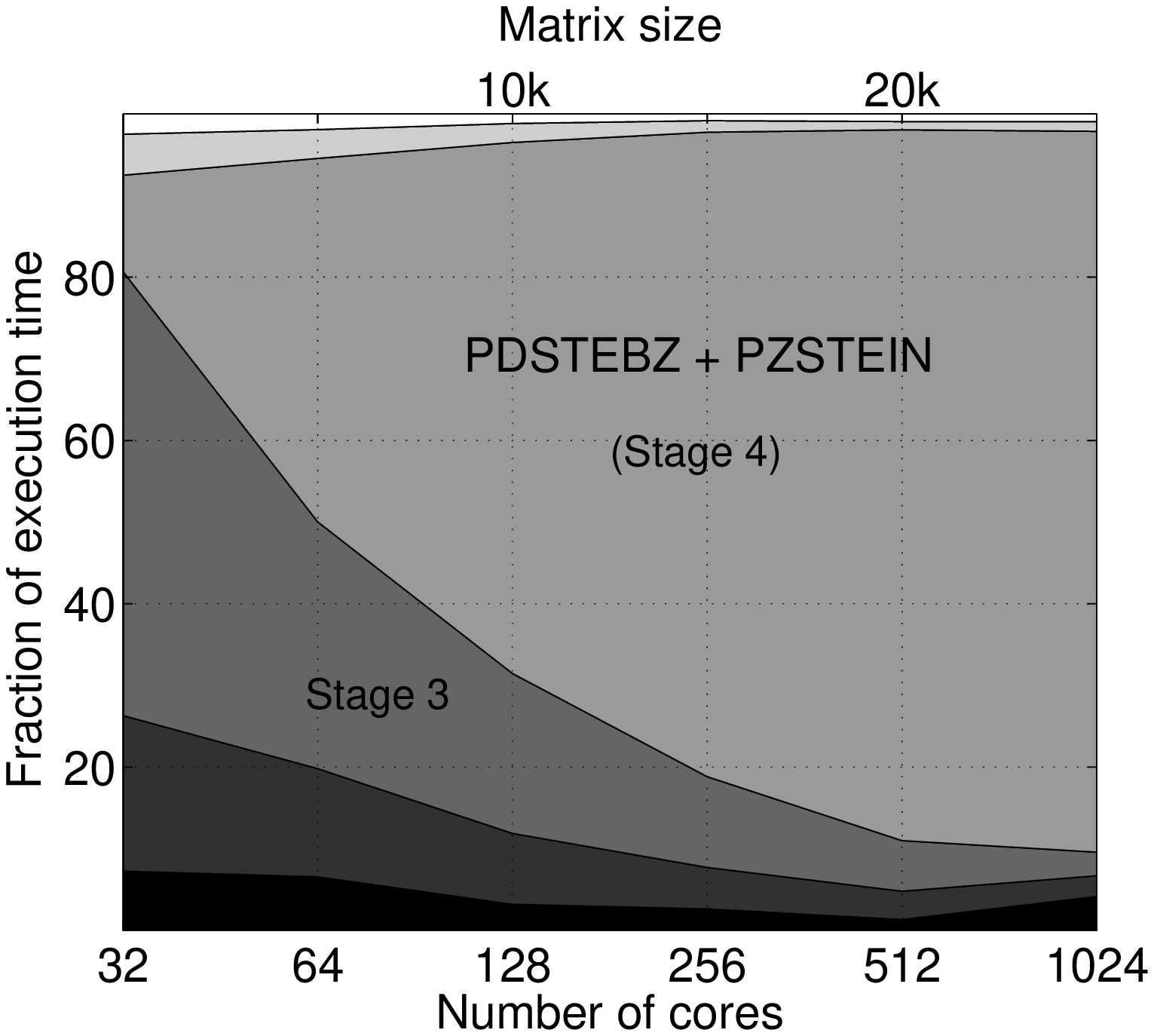}
   \caption{
     Weak scalability for the computation of 15\% of the eigenpairs
     associated with the smallest eigenvalues.
     As done commonly in practice \cite{doi:10.1021/ct900539m}, the
     eigenvectors are requested to be 
     numerically orthogonal.  
     {\it Left:}  Total execution time of {\tt PZHEGVX} and EleMRRR. 
     {\it Right:} Fraction of the execution time spent in the six stages of
     {\tt PZHEGVX},  
     from bottom to top (see Table~\ref{scalapackroutines}). 
   }
   \label{fig:timepzhegvx1}
\end{figure}

In Fig.~\ref{fig:timepzhegvx1} we report the weak
scalability of
 {\tt PZHEGVX} for computing 15\% of the eigenpairs of $Ax = \lambda
B x$ associated with
the smallest eigenvalues.\footnote{The timings were generated on the {\em Juropa} supercomputer;
a detailed description of the architecture and the experimental setup 
is provided in Section~\ref{experiments}. 
We used all default parameters.
  In particular the parameter
  {\it orfac} that indicates which eigenvectors should be re-orthogonalized
  during Inverse Iteration has the default value $10^{-3}$. 
  In practice, it is possible
  to use a less restrictive choice 
  to reduce the execution time
  to some extent. 
  In order to exploit ScaLAPACK's fastest reduction routines, the lower triangular part of the matrices 
  is stored and referenced. 
}
The left graph indicates that, as the
problem size and the number of processors increase, {\tt PZHEGVX} does
not scale as well as the EleMRRR solver presented in
Section~\ref{elemental}. In the right graph we show the breakdown of the
execution time for each of the six stages. 

Independently of the input
data, all the considered solvers perform 
a similar number of floating point operations in each of the Stages 1, 2, 3,
5, and 6; on the contrary, the complexity of Stage 4 depends on the input
data, and varies from one method to another. With the exception of
this stage, a comparison purely based on
operation counts would be misleading; differences in execution time are
mainly due to different use of the memory hierarchy and exploitation of
parallelism. Throughout the paper, we rely on the execution time as the
performance metric.


In Fig.~\ref{fig:timepzhegvx1}, it is evident that the routines {\tt
  PDSTEBZ} and {\tt PZSTEIN}, which 
implement 
the Bisection and Inverse Iteration (BI) tridiagonal eigensolver, are the main cause
for the poor performance of {\tt PZHEGVX}. For
  the problem of size $20{,}000$, these routines are responsible for
almost 90\% of the compute time.  BI's poor performance is a well
understood phenomenon, e.g.~\cite{Choi19961}, directly related to
the effort necessary to re-orthogonalize eigenvectors corresponding to
clustered eigenvalues. This issue led to the development of an
improved version of Inverse Iteration, the MRRR algorithm, that avoids
re-orthogonalization even when the eigenvalues are
clustered. In addition to the performance issue, {\tt
  PZHEGVX} also suffers from memory imbalances, as all the eigenvalues
belonging to a cluster are computed on a single processor. 


\begin{myguide}
    In light of the above considerations, the use of ScaLAPACK's
    routines based on Bisection and Inverse Iteration (BI) is not
    recommended.
\end{myguide}

We do not provide further
comparisons between EleMRRR and {\tt PZHEGVX} or {\tt PDSYGVX}.
Instead, in Section~\ref{experiments} we illustrate how the performance of
these drivers changes when the BI algorithm for the tridiagonal
eigensolver is replaced with other -- faster -- methods available in
ScaLAPACK, namely the Divide and Conquer algorithm (DC) and the MRRR
algorithm~\cite{Tisseur98parallelizingthe,Vomel:2010:SMA:1644001.1644002}.

\subsection{The Standard Eigenproblem}
\label{stdeig}

We restrict the discussion to dense Hermitian problems.
Such problems arise in a
variety of applications, ranging from electrodynamics to
macro-economics~\cite{Saad:2011}; they are also often solved as part of a
GHEP, as discussed in Section~\ref{Introduction}.

The standard approach to solve a standard eigenproblem consists of three stages,
corresponding to Stages 3--5 in the solution of the generalized problem:
(a) Reduction to a real symmetric tridiagonal form; 
(b) Solution of the tridiagonal eigenproblem;
(c) Backtransformation of the eigenvectors.
A detailed analysis of the three-stage approach, concentrating on
the first and third stage, can be found in~\cite{StanleyDiss97}.

An alternative approach
is Successive Band Reduction (SBR)~\cite{BischofSBR2000}.  
The idea is to split the reduction to tridiagonal form in two (or
more) stages. In all but the last stage, the matrix is reduced to
banded form with strictly decreasing bandwidths. 
Unlike the direct reduction to
tridiagonal form, these stages can take full advantage of highly
efficient kernels from the Basic Linear Algebra
  Subprograms (BLAS) library~\cite{blas90}, thus attaining
high-performance. 
The reduction is then completed with a final  
band-to-tridiagonal reduction.
The downside of such a strategy 
lies in the accumulation of the orthogonal transforms. Thus, when a
  significant portion of the 
eigenvectors is requested, the SBR routines are not
competitive.  For 
this reason, SBR is normally used for computing only the eigenvalues or
  a small
fraction of the eigenvectors.  However, a recent publication suggests that,
on highly 
parallel systems, SBR might be faster than direct
reduction to tridiagonal form even when a significant fraction of
eigenvectors is computed~\cite{Auckenthaler2011}.

ScaLAPACK offers a number of routines for the standard Hermitian eigenproblem,
each of which differs in its algorithm of choice for the tridiagonal
eigenvalue problem. 
The four routines {\tt PZHEEVX}, {\tt PZHEEV},
{\tt PZHEEVD}, and the recently added {\tt PZHEEVR}, implement
BI~\cite{wilkinson58}, the QR 
algorithm~\cite{qr61b,qr61a}, the 
DC algorithm~\cite{dc81,dc95}, and the MRRR
algorithm~\cite{Dhillon04multiplerepresentations},
respectively.\footnote{Each of these routines has the usual counterpart for
  the real symmetric case: {\tt PDSYEVX}, {\tt PDSYEV},
{\tt PDSYEVD}, and {\tt PDSYEVR}.}

Among the solvers presently available in ScaLAPACK, only {\tt PZHEEVX} (BI) and 
{\tt PZHEEVR} (MRRR) offer the possibility of computing a subset of
eigenpairs. {\tt PZHEEVX} is widely used, even though, as highlighted in
the previous section, it is highly non-scalable. 
Similarly, if eigenvectors are computed, the QR algorithm is known to be
slower than DC for large problems~\cite{Bientinesi:2005:PED:1081198.1081222}
and thus {\it the use of 
  ScaLAPACK's routines based on QR is not recommended}; in the future
experiments, we omit 
comparisons with routines that are based on the QR algorithm or BI.

\begin{figure}[!thb]
   \centering
\includegraphics[scale=.38]{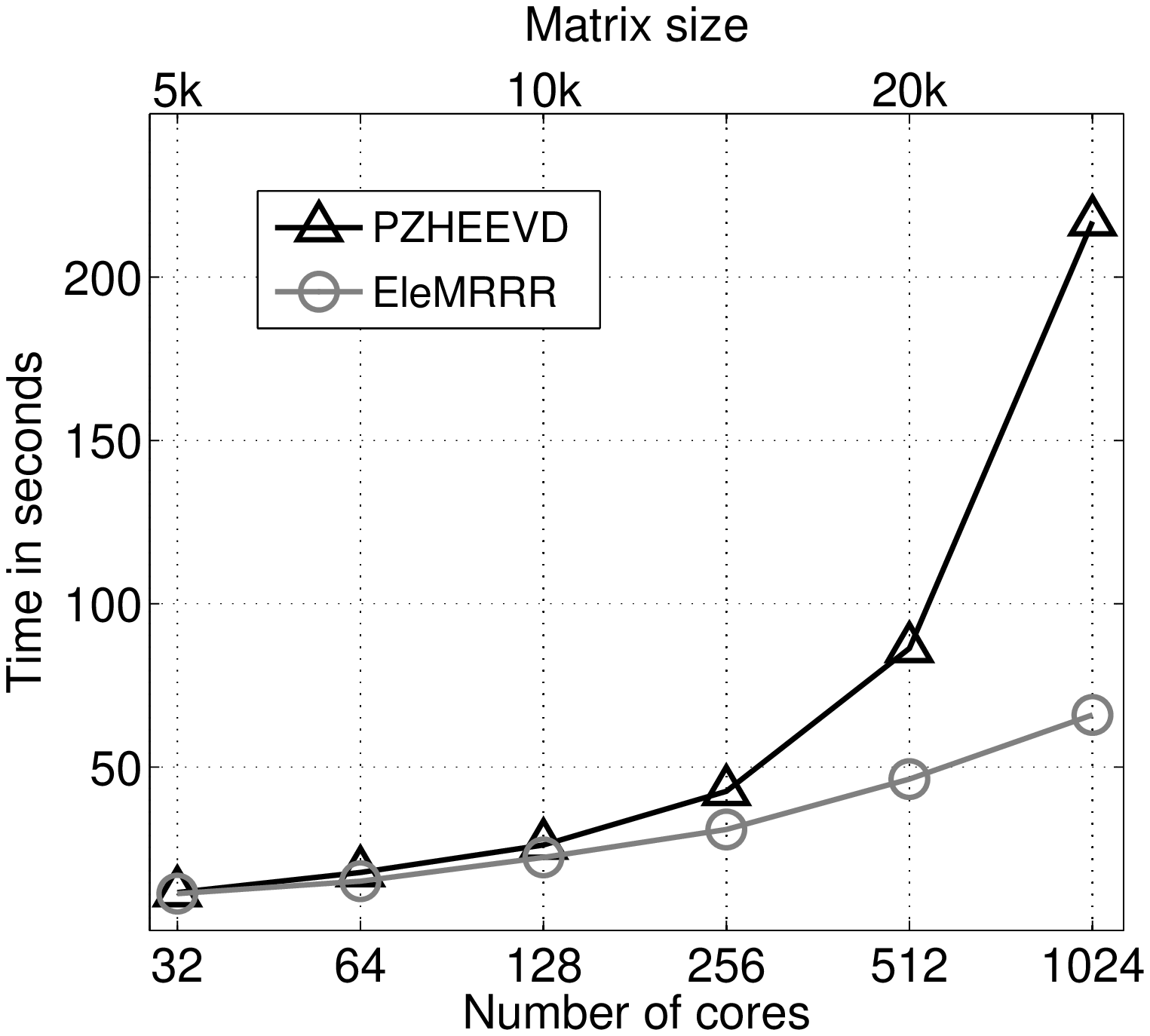}
\includegraphics[scale=.38]{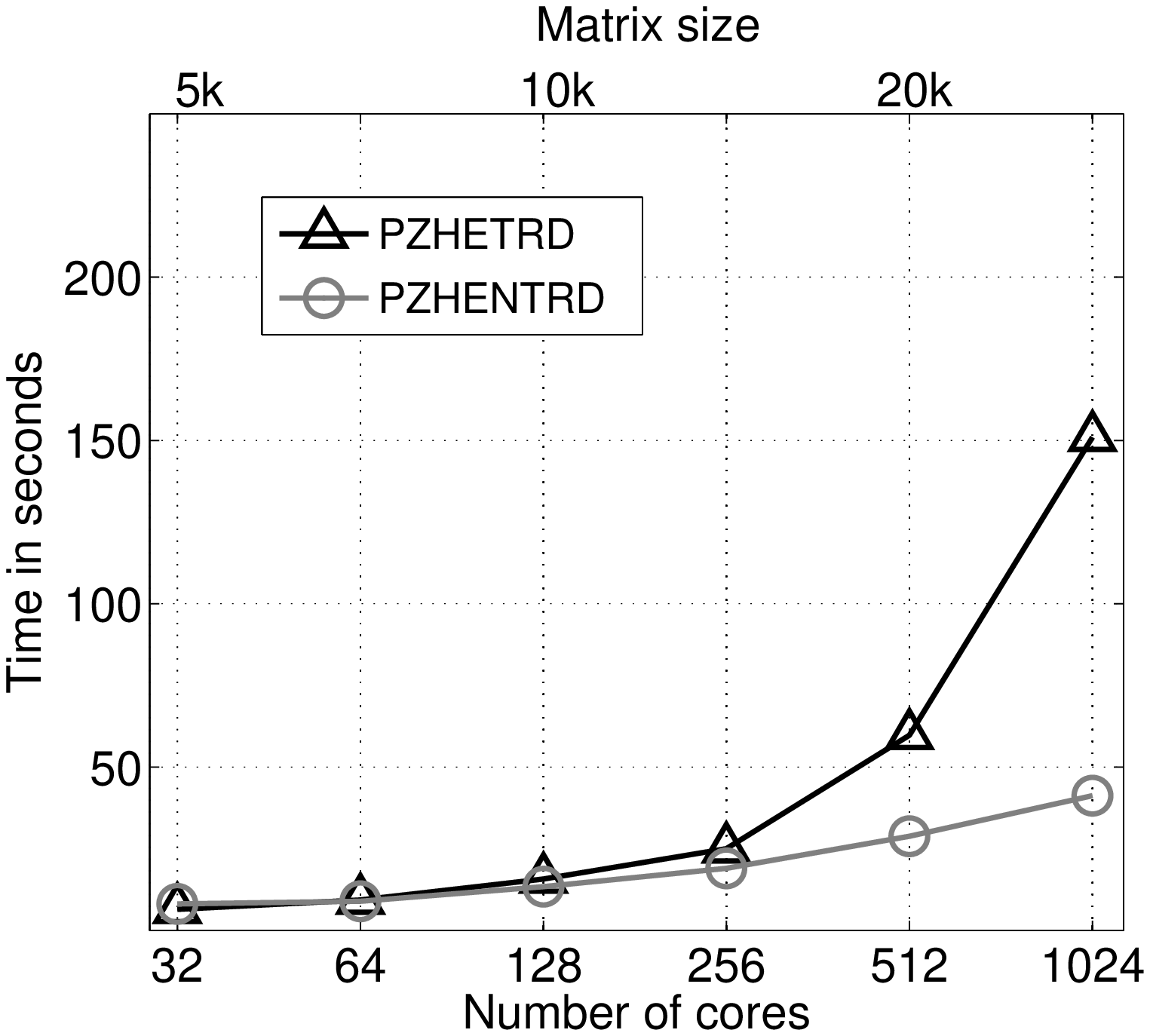}
   \caption{
     Weak scalability for the computation of all eigenpairs using DC.
     {\it Left:}  Total execution time of {\tt PZHEEVD} and EleMRRR.
     {\it Right:} Execution time for ScaLAPACK's routines {\tt PZHETRD} and
     {\tt PZHENTRD}, which are responsible for the reduction to tridiagonal
     form. The former,
     used within the routine {\tt PZHEEVD}, 
     causes a performance penalty and accounts for much of the time
     difference compared with EleMRRR.
   }
   \label{fig:timepzheevd2}
\end{figure}

We now focus on the weak scalability of the widely used routine, {\tt
  PZHEEVD}, which uses DC for the tridiagonal eigenproblem. 
In Fig.~\ref{fig:timepzheevd2}, we show the results for {\tt PZHEEVD} and
EleMRRR from an experiment similar to that of Fig.~\ref{fig:timepzhegvx1}.
Note that all eigenpairs were computed, since {\tt PZHEEVD} does not allow
for subset computation.
In the previous section, the example indicated that BI might dominate the
runtime of the entire dense eigenproblem, while the DC method required less
than 10\% of the total execution time. 
Instead, as the matrix size increases, the reduction to tridiagonal
form ({\tt PZHETRD}) becomes the computational bottleneck, requiring
up to 70\% of the total time for the largest problem shown. A comparison of
the left and right sides of Fig.~\ref{fig:timepzheevd2} 
reveals that, for large problems, using {\tt PZHETRD} for the reduction to
tridiagonal form requires more time than the complete solution with
EleMRRR. 

ScaLAPACK also includes {\tt PZHENTRD}, a routine for the reduction to
STEP especially optimized for square processor grids. The performance
improvement with respect to {\tt PZHETRD} can be 
so dramatic that, for this stage, it is
  preferable
to limit the computation to a square number of
processors and redistribute the matrix
accordingly~\cite{Hendrickson1999}. It is important to note
that the performance benefit of {\tt PZHENTRD} can only be exploited if the
lower triangle of the input matrix is stored,
otherwise the slower routine, {\tt PZHETRD}, is invoked.\footnote{Similar
  considerations also apply for the reduction to  
  standard form via the routines {\tt PZHEGST} and {\tt PZHENGST},
  see~\cite{Elemental2:TOMS}.}  
%
%

\begin{myguide}
ScaLAPACK's reduction routines ({\tt PxxxNGST} and {\tt
      PxxxNTRD}) optimized for square grids of processors are to be preferred
    over the regular reduction 
  routines, even when non-square process grids are used;
  moreover, only the lower triangle of implicitly Hermitian matrices should
  be referenced. 
\end{myguide}


For performance and scalability reasons, in Section~\ref{experiments} we
 use the  
routines {\tt PxxxNTRD} and {\tt PxxxNGST} 
to build the fastest solver within the ScaLAPACK framework.

\subsection{The Tridiagonal Eigenproblem} 
\label{trdeig}

At the core of the reduction-backtrans\-formation approach for the GHEP and
the HEP is the symmetric tridiagonal eigenproblem.  
One of the main
differences -- and often the only difference -- among solvers for
generalized and 
standard problems lies in the method for this stage.  
As seen in Section~\ref{geneig}, this might account for a significant
computational portion of the entire solution process.  
In Section~\ref{stdeig}, 
we mentioned four methods: BI, QR, DC, and MRRR, and justified not providing
experimental comparisons with the BI and QR approaches. 

\begin{figure}[thb]
   \centering
   \includegraphics[scale=.38]{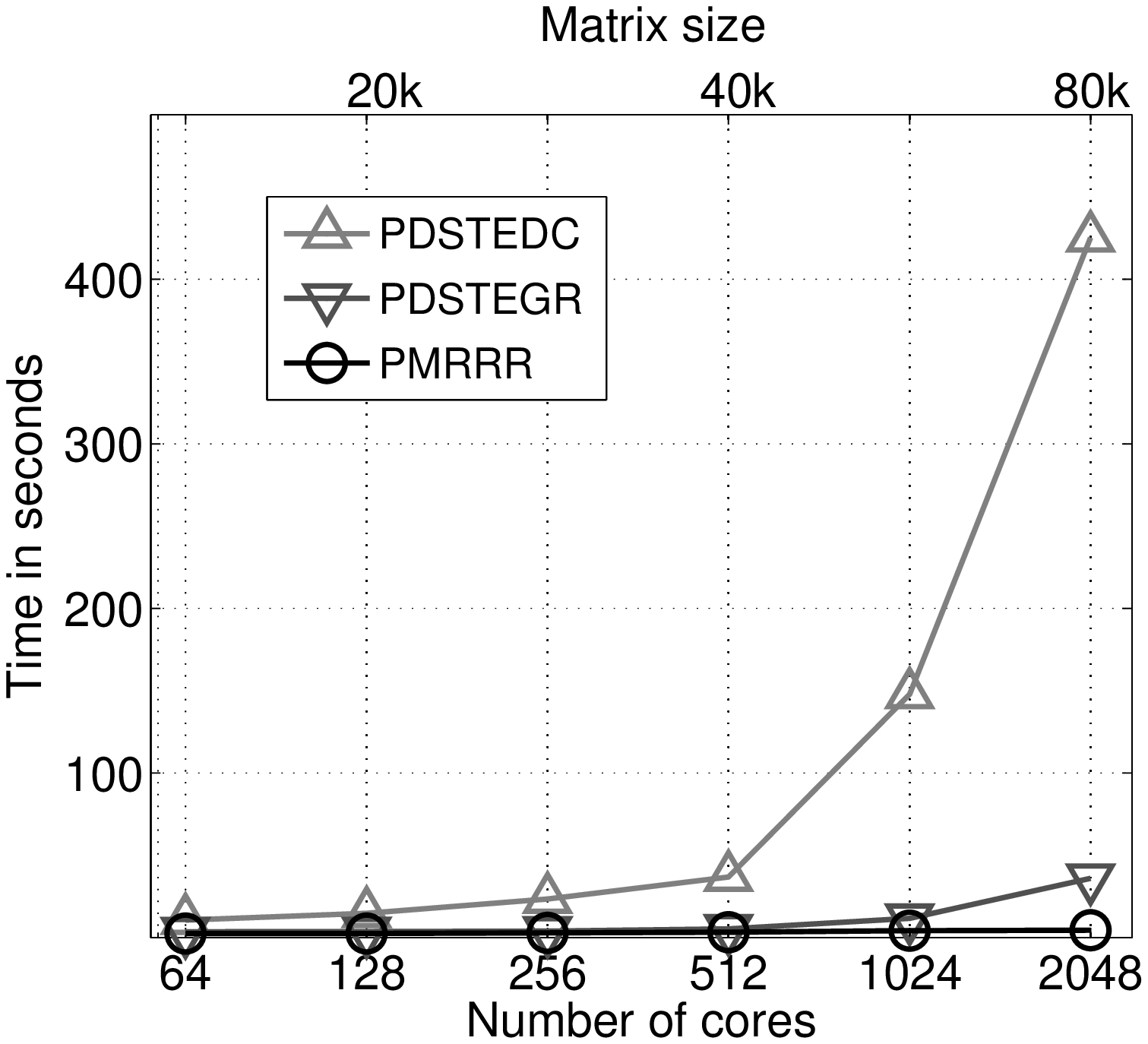} 
\includegraphics[scale=.38]{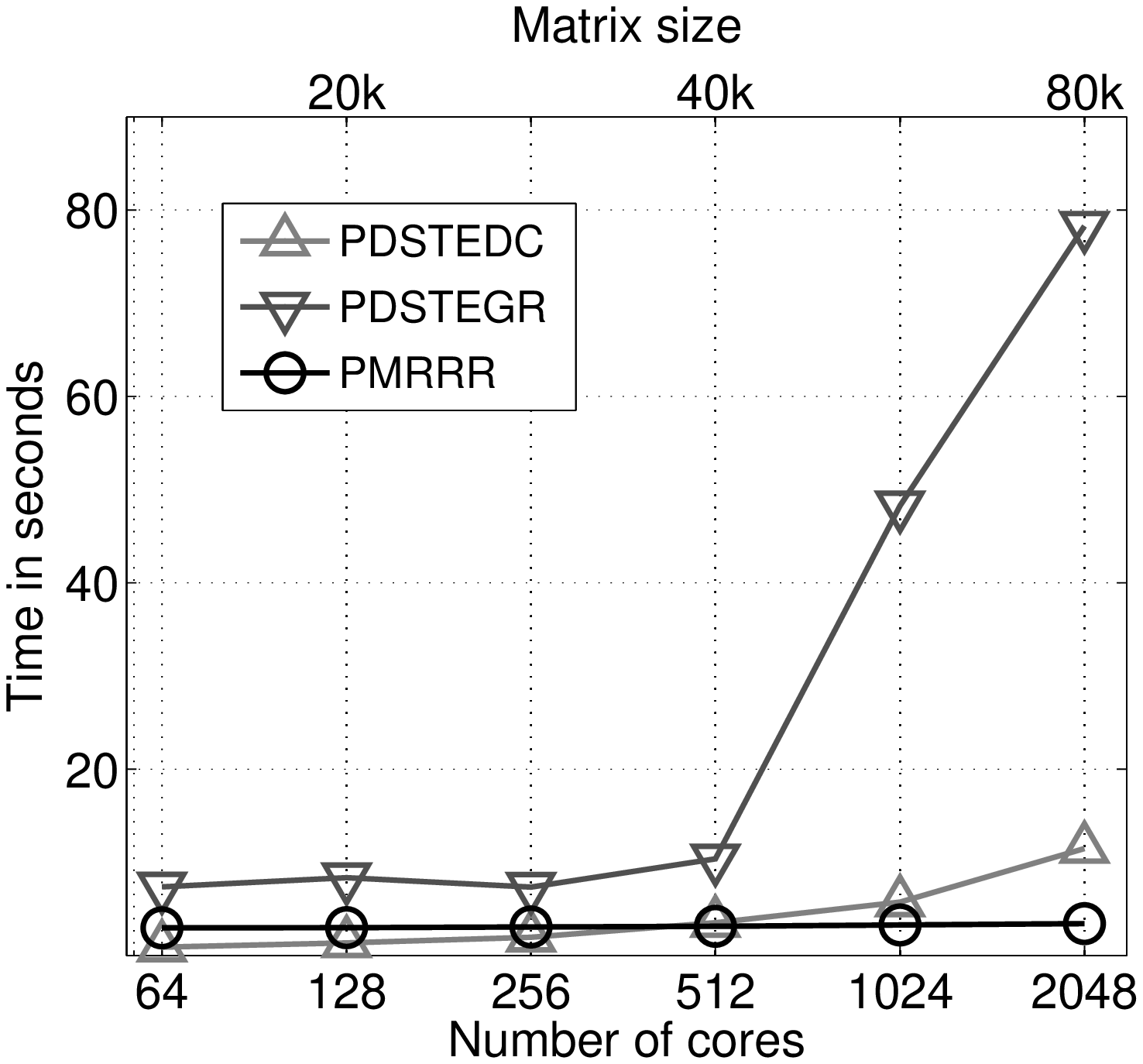}
   \caption{
     Weak scalability for the computation of all eigenpairs of 
     two different test matrix types. The left and right graphs have
     different scales. 
     {\it Left:} ``1--2--1'' matrix; 
     {\it Right:} ``Wilkinson'' matrix.
     In contrast to the results reported
     in~\cite{Vomel:2010:SMA:1644001.1644002}, where a similar experiment
     comparing {\tt PDSTEDC} and {\tt
  PDSTEGR} is performed,  
     even when the matrices offer an opportunity for heavy deflation,
     our PMRRR becomes faster than DC eventually due to its superior
       scalability. The scalability advantage of PMRRR compared with {\tt
  PDSTEGR} is the result of non-blocking communications used in conjunction
with a task-based algorithmic design that allows processes to 
continue computing while waiting for data. 
As an example, in the Wilkinson experiment with 1024 cores, ScaLAPACK's MRRR
spends 
about 30 out of 50 seconds in exposed communication.
   }
   \label{fig:timetrdeig}
\end{figure}

As already mentioned, in contrast to all other stages of generalized and
standard problems, the number 
of arithmetic operations of the tridiagonal eigensolver depends on the input
data. In fact, depending on the  
matrix entries, either DC or MRRR may be faster.
Fig.~\ref{fig:timetrdeig} provides an example of how performance is
influenced by the input data. 
The algorithms are compared on two types of test matrices:
``1--2--1'' and ``Wilkinson''.
The former contains ones on the subdiagonals
and twos on the diagonal; its eigenpairs are known analytically.
In the latter, the subdiagonals contain ones and the 
diagonal equals the vector 
$(m, m-1, \dots, 1, 0, 1, \dots, m)$, with $m = (n-1)/2$. 
Due to the phenomenon of deflation, this matrix is known to favor the DC
algorithm~\cite{dc81}. 
We also include timings for our solver, PMRRR; for both matrix types it
eventually becomes the fastest solver.
A detailed discussion of PMRRR follows in the next section. 

We now provide a short comparison of PMRRR with ScaLAPACK's routines {\tt
  PDSTEDC} and {\tt PDSTEGR}, which implement the tridiagonal DC and MRRR
algorithms\footnote{{\tt
  PDSTEGR} is not part of ScaLAPACK; it corresponds to the tridiagonal
  eigensolver used in {\tt
  PxxxEVR} and is not encapsulated in its own routine.},
respectively~\cite{Tisseur98parallelizingthe,Vomel:2010:SMA:1644001.1644002}. 

In the worst case, {\tt PDSTEDC} computes all eigenpairs at the cost
of $O(n^3)$ floating point operations ({\em flops}); in practice, the
flop count is lower due to deflation and the run time behavior
  is empirically $n^{2.5}$~\cite{perf09,Tisseur98parallelizingthe}. 
Furthermore, most of the computation is cast in terms of fast BLAS-3
kernels~\cite{perf09}. 

 While {\tt PDSTEDC} cannot compute a subset of
$k < n$ eigenpairs, the MRRR routine {\tt PDSTEGR} returns the
eigenpairs at the reduced cost of $O(nk)$ flops. 





\begin{myguide}
Both of ScaLAPACK's tridiagonal eigensolver routines,
    {\tt
      PDSTEDC} (DC) and {\tt PDSTEGR} (MRRR), are generally 
    fast and reasonably scalable; depending on the target architecture and
    specific requirements from the applications, either one may be
    used. Specifically, if only a small 
subset of the spectrum has to be computed, 
in terms of performance, the MRRR-based solvers are to be
preferred to DC.
\end{myguide}

In terms of accuracy, {\em all} tridiagonal solvers generally obtain
accurate results in the sense that they achieve small {\em residual norms} and
{\em numerical 
orthogonality}: Specifically, with machine precision $\varepsilon$,
\begin{align}
  \| T \hat{z}_j - \hat{\lambda}_j \hat{z}_j\| &= \mathcal{O}(n \varepsilon
  \| T \|) \,, \mbox{and} \label{Eq:Tresnorm} \\ 
  |\hat{z}_j^T \hat{z}_i| &=
  \mathcal{O}(n \varepsilon), \; i \neq j \,, \label{Eq:Torthogo}
\end{align}
for all computed eigenpairs $(\hat{\lambda}_j, \hat{z}_j)$ with $\|
\hat{z}_j \|_2 \approx 1$.
It is well-known that MRRR yields slightly worse coefficients
for~Eq.~\eqref{Eq:Torthogo} than QR and DC~\cite{perf09}.
In the rare cases where this is an issue, feasible alternatives are the DC
algorithm and a mixed precision variant of MRRR~\cite{mr3smpmixed}. 
On the downside, DC requires $O(n^2)$ extra memory;
see~\cite{perf09,Bientinesi:2005:PED:1081198.1081222,Vomel:2010:SMA:1644001.1644002,mr3smpmixed} for further comparisons. 

\begin{figure}[!thb]
   \centering
   \includegraphics[scale=.38]{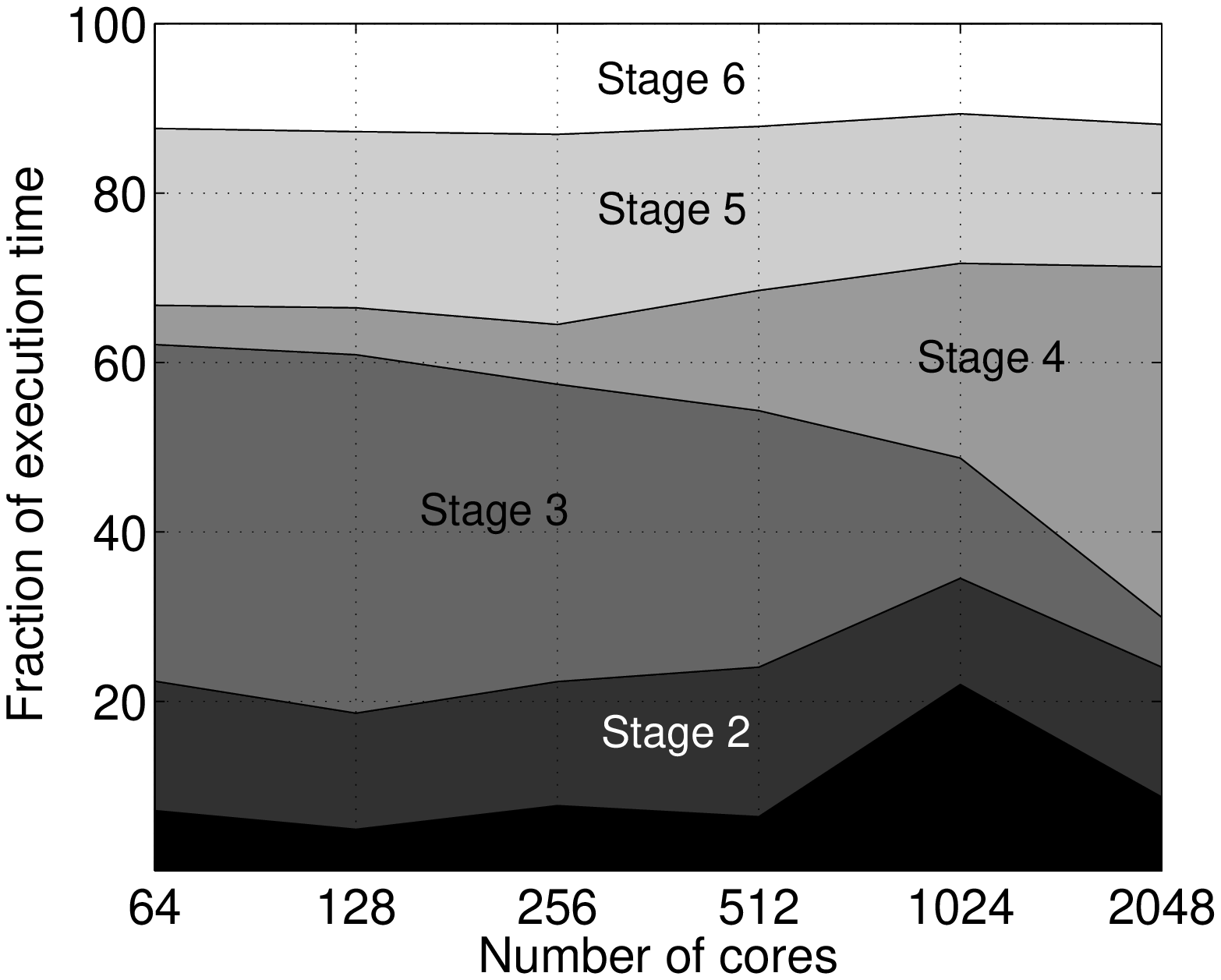} 
\includegraphics[scale=.38]{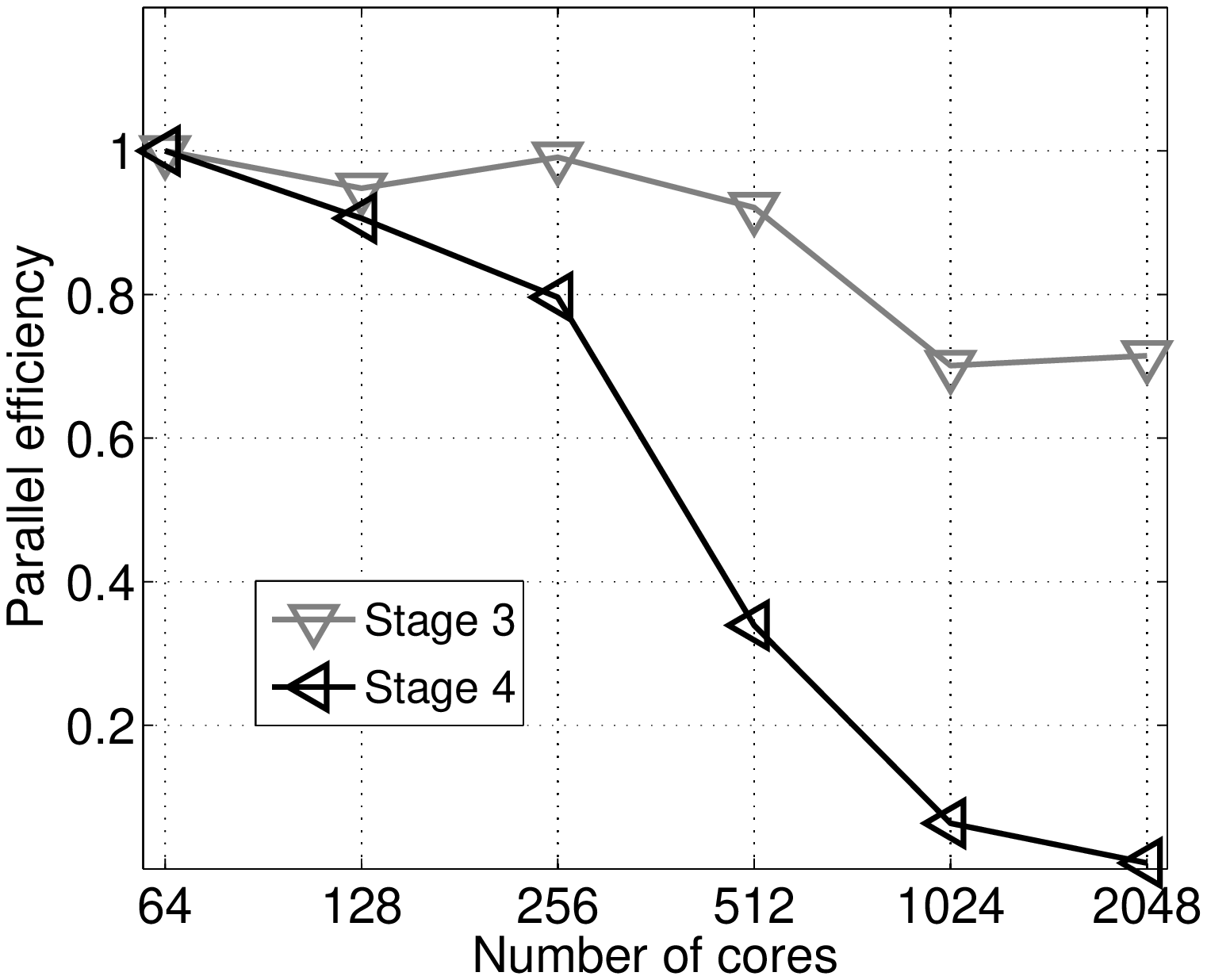}
   \caption{
     Scalability of the computation of all eigenpairs using ScaLAPACK's DC.
     The details of the experiment are discussed in
     Section~\ref{sec:strngscal}.
   }
   \label{fig:strngscaldcfrac}
\end{figure}

In Section~\ref{experiments}, we include experimental data on generalized
eigenproblems for both DC and MRRR.  
One of the challenges in building a scalable solver is that every stage must
be scalable.  
This is illustrated in
Fig.~\ref{fig:strngscaldcfrac}, which shows the results for ScaLAPACK's DC
from the experiment detailed in Section~\ref{sec:strngscal}. 
While the reduction to tridiagonal form
and the tridiagonal eigensolver
are respectively the most and the least expensive stages on 64 cores,
on 2048 cores the situation is reversed.
This behavior is explained by the parallel efficiencies shown in the right
panel of Fig.~\ref{fig:strngscaldcfrac}. 

\section{Elemental's Generalized and Standard Eigensolvers}
\label{elemental}

Elemental is a framework for dense matrix computations on distributed-memory
architectures~\cite{ElemTOMS}. The main objective of the
project is to ease the process of implementing matrix operations
without conceding performance or scalability.
The code resembles a high-level description of the algorithm, and
the details relative to data distribution and communication are hidden.
Performance is achieved through a two-dimensional cyclic elemental (entry-wise) 
matrix distribution.\footnote{ScaLAPACK instead deploys a block-cyclic matrix distribution.}

Elemental supports many
frequently encountered dense matrix operations; these include the
Cholesky, QR and LU factorizations, as well as the solution of linear
  systems.  Recently, 
routines for Hermitian and symmetric eigenvalue problems have also
been added.  In particular, Elemental supports Hermitian-definite
generalized, standard Hermitian, as well
as skew-Hermitian eigenproblems.
All the eigensolvers follow the classical reduction and
backtransformation approach described in Section~\ref{Introduction}.
Detailed discussions the reduction and backtransformation stages, both
in general and within the Elemental environment, can be found
in~\cite{StanleyDiss97,Sears:1998,Hendrickson1999,Elemental2:TOMS}.


In terms of accuracy, Elemental's solvers are equivalent to their sequential
counterparts. We therefore do not report accuracy results and refer to
\cite{perf09} instead. 

In terms of memory, Elemental's solvers are quite efficient. In
Table~\ref{tab:memoryusgae}, we report approximate total memory
requirements for computing $k$ eigenpairs of generalized and standard
eigenproblems. As 
a reference, we provide the same numbers for the solvers built from ScaLAPACK
routines. In Section~\ref{experiments}, we define the meaning of ``ScaLAPACK
DC'' and ``ScaLAPACK 
MRRR'' precisely.  The numbers in
the table are expressed in units of the size of a 
single complex and real floating point number, depending on the required
arithmetic. The memory requirement per process can be obtained by dividing
by the total number of processes.

\begin{table}[htb]
\begin{center}
\footnotesize
\begin{tabular}{ c c c c c} \hline\noalign{\smallskip}
    & \multicolumn{2}{c}{Complex}        & \multicolumn{2}{c}{Real} \\ 
    & GHEP             & HEP             & GHEP         &  HEP \\
\hline\hline\noalign{\smallskip}
      ScaLAPACK DC    & $4 n^2$          & $3 n^2$  & $5 n^2$        & $4 n^2$  \\[1mm]
      ScaLAPACK MRRR  & $2 n^2 + 1.5 nk$ & $n^2 + 1.5nk$  & $2 n^2 + 2 nk$ &
      $n^2 + 2nk$  \\[1mm]
      Elemental         & $2 n^2 + nk$     & $n^2 + nk$  & $2 n^2 + nk$   & $n^2 + nk$ \\[1mm]
  \hline 
\end{tabular}
\end{center}
\caption{Approximate total memory requirements in units of complex and real
  floating point numbers for the computation of $k$ eigenpairs. 
  We assume that square process grids
  are used, as it is recommended for optimal 
performance {\it and} if memory usage is a concern.  
  The numbers also hold when a non-square grid of processes is used in
  combination with the standard
  reduction routines.
}
\label{tab:memoryusgae}
\end{table}

For square process grids, Elemental requires between $0.5n^2$ to $2n^2$
floating point 
numbers less memory than ScaLAPACK based solvers. 
If non-square grids are
used, a user concerned about memory usage can make use of the non-square
reduction routines -- at the cost of sub-optimal performance. The
  reduction routines for square process grids would otherwise need a data
  redistribution that adds a $n^2$ term to
Elemental and ScaLAPACK MRRR, but not to DC. On the other hand, in this
situation, ScaLAPACK
MRRR can save work space to perform its redistribution of the eigenvectors
from a 1d to a 2d block-cyclic data layout, reducing its terms by $nk$ real
floating point numbers. This eigenvector redistribution is performed in
Elemental in-place, resulting in a smaller memory footprint than possible for
ScaLAPACK's routines. 

For the solution of symmetric tridiagonal eigenproblems, Elemental
incorporates PMRRR, a new 
parallel variant of the MRRR algorithm.  PMRRR combines two solvers:
one based on message-passing
parallelism~\cite{Bientinesi:2005:PED:1081198.1081222,Vomel:2010:SMA:1644001.1644002},
and another based on multithreading~\cite{mr3smp}.
As a consequence, the solver
provides the user with 
multiple parallel programming models 
(multithreading, message-passing, a hybrid of both),
so that 
the parallelism offered by modern distributed-memory architectures
can be fully exploited. In
the following two subsections we focus the attention on PMRRR, our
tridiagonal eigensolver.

\subsection{PMRRR as the Core of Elemental's Eigensolver}


The MRRR algorithm is a form of inverse iteration. 
Its salient feature is that 
the costly re-orthogonalization, necessary when 
the eigenvalues are clustered together, is entirely removed.
As the experiment described in Fig.~\ref{fig:timepzhegvx1} shows, 
the re-orthogonalization may affect the execution time dramatically:
in computing 15\% of the eigenpairs associated with the smallest eigenvalues  
of a matrix of size $20{,}000$ using 512 cores, 
ScaLAPACK's Inverse Iteration takes about 404 seconds, while
PMRRR requires less than $0.3$ seconds. 

To achieve this performance, 
the classical inverse iteration 
was modified significantly. The basic procedure requires the
repeated solution of linear systems: given an approximate eigenvalue 
$\hat{\lambda}_j$ and a starting vector $\hat{z}_{j}^{(0)}$, under mild assumptions\footnote{In particular,  
    $|\lambda_j - \hat{\lambda}_j| < \max_{i \neq j} |\lambda_i -
    \hat{\lambda}_j|$ is required.} the iterative
process 
\begin{equation}
(T - \hat{\lambda}_j I) \hat{z}_{j}^{(i+1)} = s^{(i)} \hat{z}_{j}^{(i)} \,,
\end{equation}
with appropriate scaling factors $s^{(i)}$, results in an eigenvector
approximation 
$\hat{z}_{j}$ that fulfills
Eq.~\eqref{Eq:Tresnorm}. Nevertheless, small
residual norms do not 
guarantee orthogonality between independently computed eigenvectors -- that is
Eq.~\eqref{Eq:Torthogo} 
might not hold~\cite{Ipsen97computingan}. 

The close connection between inverse 
iteration and the MRRR algorithm is discussed in a number of
articles~\cite{NLA:NLA493,Dhillon:2006:DIM:1186785.1186788}.
Here we only mention three differences:
(1) Instead of representing tridiagonal matrices by their diagonal and
subdiagonal entries, MRRR uses {\em Relatively Robust Representations}
(RRRs); these are representations with the property that small
relative perturbations to the data result in small relative
perturbations to the eigenvalues~\cite{Parlett2000121}; 
(2) the selection of a nearly optimal right-hand side vector
$\hat{z}_{j}^{(0)}$ for a 
one-step inverse iteration~\cite{Fernando97}; 
(3) the use of so-called {\em twisted factorizations} 
 to solve linear systems~\cite{Fernando97}.  

 In the following, we briefly discuss both the mathematical foundations
 of the MRRR algorithm, and how parallelism is organized in PMRRR.
 
\subsubsection{The MRRR algorithm} 
At first, a representation {\it RRR$_{\it 0}$} is chosen so that it defines
all the desired eigenpairs in a relatively robust
way~\cite{Dhillon04multiplerepresentations,WillemsDiss10}.\footnote{
Without loss of generality, we will assume that
$T$ is numerically irreducible: No off-diagonal element is smaller
in magnitude than a certain threshold. In context of the dense
  eigenproblem, the threshold is usually set to $\varepsilon \|T\|$.} 
For example, a factorization of the form 
$L_0 D_0 L_0^T = T - \sigma I$ is an RRR for all of the
eigenvalues~\cite{Parlett2000121}; 
here, $D_0$ is diagonal, $L_0$ is lower unit bidiagonal, 
and $\sigma$ is a scalar such that $T - \sigma I$ is definite.
As a second step, bisection is used to compute
approximations to each  
eigenvalue $\hat{\lambda}_j$. 
At the cost of $\mathcal{O}(n)$ flops, bisection
guarantees that each  
eigenvalue is computed with high relative
accuracy~\cite{Marques:2006:BIF}.

At this point,
for all the eigenvalues $\hat{\lambda}_j$ that have large\footnote{In
  practice, a relative gap greater than $10^{-3}$ is 
  considered sufficiently large.} relative
gaps, i.e., $\mbox{relgap}( \hat{\lambda}_j) = \min_{i \neq j}
|\hat{\lambda}_j - \lambda_i| / |\hat{\lambda}_j| > \mbox{tol}$, 
it is possible to compute their corresponding eigenvectors: 
By using one step of inverse iteration with a twisted factorization,
the acute angle $\angle (\hat{z}_j,z_j)$ between the computed eigenvector
$\hat{z}_j$ and true eigenvector $z_j$ satisfies
\begin{equation}
  \label{eq:errorangle}
  |\sin \angle (\hat{z}_j,z_j)| \leq \mathcal{O}\left( \frac{n \varepsilon}{\mbox{relgap}(
    \hat{\lambda}_j) } \right) \enspace \,.
\end{equation}
The denominator in Eq.~\eqref{eq:errorangle} is the reason for which 
the process can be applied only to well-separated eigenvalues -- measured by
its relative gap. Such
well-separated eigenvalues are called {\it singletons}. Indeed, for
singletons the 
computed eigenvectors have a small angle to the  
true eigenvector and consequently are numerically orthogonal. 

In contrast, when a set of consecutive eigenvalues have small relative
gaps -- that is, they form a {\it cluster} -- the current RRR cannot
be used to obtain numerically orthogonal eigenvectors.  Instead, one
exploits the fact that the relative gap is not invariant to matrix
shifts.  The algorithm then proceeds by constructing a new RRR $\{L_c,
D_c\}$ in a mixed relatively stable way: $L_c D_c L_c^T = L_0 D_0
L_0^T - \sigma_c I$. 
By shifting, the relative gaps are 
modified by a factor
$|\hat{\lambda}_j|/|\hat{\lambda}_j - \sigma_c|$;
thus $\sigma_c$ is chosen close to one of the eigenvalues in the cluster,
so that at least one of them becomes well-separated.  
Once the new RRR is established, the eigenvalues in the cluster are refined 
to high relative accuracy with respect to the new RRR and classified as
singletons and clusters. 
If all eigenvalues in the cluster are singletons, then the
  representation $\{L_c, D_c\}$ can be used to compute a set of 
  orthogonal eigenvectors for a slightly perturbed invariant subspace of $\{L_0,
  D_0\}$. If instead some eigenvalues are still clustered, the procedure is
  repeated recursively until all desired eigenpairs are computed.

    As a detailed discussion of the MRRR algorithm is outside the scope of
    this article, for further information we refer the readers
    to~\cite{Fernando97,DhillonDiss97,ortvecs04,Dhillon04multiplerepresentations,WillemsDiss10}
    and the references therein.

\subsubsection{PMRRR's parallelism}
Our parallelization strategy consists of two layers: a global and a local
one. At the global level, the $k$ eigenvalues and eigenvectors are
  statically divided
  into equal parts and assigned to the processes. Since the unfolding of the
  algorithm depends on the spectrum, it is still possible that the workload
  is not perfectly balanced among the processes. At the local level (within each
  process), the computation is decomposed into tasks that can be executed in
  parallel by multiple threads. The processing of these tasks leads to the
  dynamic generation of new tasks and might involve communication with other
  processes. The newly generated tasks are then likewise enqueued. 

When executed with $nproc$ processes, the algorithm starts by
broadcasting the input matrix and by redundantly computing the initial
representation {\it RRR$_{\it 0}$}. Once this is available, the computation
of the approximations $\hat{\lambda}_j$ of $k$ eigenvalues of $L_0 D_0 L_0^T$
is embarrassingly parallel: Each process is responsible for at most
$epp = \lceil k/nproc \rceil$ eigenvalues;\footnote{PMRRR (ver.~0.6)
  computes approximations to all $n$ eigenvalues at this stage, such that
  $epp = \lceil n/nproc \rceil$.}
 similarly, each of the
$nthread$ threads within a process 
has the task to compute at most $ept = \lceil
epp/nthread \rceil$ eigenvalues.  The processes then gather all the
eigenvalues, and the corresponding eigenpairs are assigned 
as desired. 

Locally, the calculation of the eigenvectors\footnote{We only
  refer to the calculation of the {\it eigenvectors} here, although the
  eigenvalues are also modified in this part of the computation.}  is split
into 
computational tasks of three types: a set of singletons, clusters that
require no 
communication, and clusters that require 
communication with other processes.
The computation associated with each of the three types is detailed below. 
  \begin{enumerate}
    \item {\it A set of singletons.} The corresponding
      eigenvectors 
      are computed locally. No further communication among
      processes is necessary.
    \item {\it A cluster requiring no communication.} When the cluster 
      contains eigenvalues assigned to only one process, no
      cooperation among processes  
      is needed. The 
      four necessary steps are the same as those for the cluster task
      in~\cite{mr3smp}: 
      A new RRR is computed; 
      the eigenvalues are refined to relative accuracy with respect to the
      new RRR; the refined eigenvalues are 
      classified into singletons and clusters;
      the corresponding tasks are enqueued into the local work queue. 
    \item {\it A cluster requiring communication.} 
      When the cluster contains a set of eigenvalues which spans multiple processes,
      inter-process communication is needed.
      In this case, all the processes involved perform the following steps:
      a new RRR is computed redundantly, the local set of eigenvalues is
      refined, and the eigenvalues of the cluster are gathered and
      reclassified. 
    \end{enumerate}

    Multithreading support is easily obtained by
    having multiple threads dequeue and execute tasks. The tasks are
    dynamically generated: their number and size highly depends on the
    spectral distribution of the input matrix; for this reason,
    the execution time for matrices of the same size may differ
    noticeably. By contrast,  
    the memory requirement is matrix independent ($\mathcal{O}(nk/nproc)$
    floating point numbers per
    process), 
    and perfect memory balance is achieved~\cite{Bientinesi:2005:PED:1081198.1081222,Vomel:2010:SMA:1644001.1644002}.

Our tests show that the hybrid parallelization approach is equally or slightly
faster than the one purely based on MPI. This is generally true for
architectures with a high degree of inter-node parallelism and limited
intra-node parallelism. 
By contrast, on architectures with a small degree
of inter-node parallelism and high degree of intra-node parallelism, we
expect the hybrid execution of 
PMRRR to be preferable to pure MPI.

  We stress that even when no multithreading is used, the task-based
  design of PMRRR can be advantageous: By scheduling tasks that require
  inter-process communication with priority and using non-blocking
  communication, processes can continue executing tasks while waiting to
  receive data. This strategy often leads to a perfect overlap of
  communication and computation.

\section{Experiments}
\label{experiments}



In the next two sections we present experimental results for the execution
 on two state-of-art supercomputers at the
Research Center J\"ulich, Germany: {\em Juropa} and {\em Jugene}.







\subsection{Juropa}
\label{juropa}




Juropa consists of 2208 nodes, each comprising two Intel
{\it Xeon X5570 Nehalem} quad-core processors running at 2.93 GHz
with 24 GB of memory. The nodes are connected by an {\it Infiniband QDR}
network with a Fat-tree topology.

All tested routines were compiled using the {\it Intel compilers}
(ver.~11.1) with the flag {\tt -O3} and linked to the {\it
  ParTec's ParaStation MPI} 
library 
(ver.~5.0.23).\footnote{Version 5.0.24 was used when support for
  multithreading was needed.} 
Generally, we used a two-dimensional
process grid $P_r \times P_c$ (number of rows $\times$ number of columns)
with $P_r = P_c$ whenever possible, and 
$P_c = 2P_r$ otherwise\footnote{As discussed in Sections~\ref{stdeig} and
  \ref{elemental}, $P_c
  \approx P_r$ or the 
  largest square grid possible should be
  preferred. These choices do
  not affect the qualitative behavior of our performance results.}. 
If not
stated otherwise, one process per core was employed.

The ScaLAPACK library (ver.~1.8) was used in conjunction with
Intel's MKL BLAS (ver.~10.2).\footnote{At the time of writing, ScaLAPACK's
  latest version was 1.8. The current version, 2.0, presents no
  significant changes in the tested routines.}
From extensive testing, we identified that in
all cases the optimal block size was close to 32;
therefore we carried out the ScaLAPACK experiments only with
block size of 16, 32, 48; the best result
out of this pool is then reported.

Since no driver for the generalized eigenproblem that makes use of DC is
available,   
we refer to ScaLAPACK's DC as the sequence of routines 
{\tt PZPOTRF}--{\tt PZHENGST}--{\tt PZHENTRD}--{\tt PDSTEDC}--{\tt
  PZUNMTR}--{\tt PZTRSM}, as listed in
Table~\ref{scalapackroutines}. 
Similarly, we refer to ScaLAPACK's MRRR as the same 
sequence with {\tt PDSTEDC} replaced by {\tt PDSTEGR}.\footnote{As {\tt
    PDSTEGR} is not contained in ScaLAPACK, it corresponds to the sequence {\tt PZPOTRF}--{\tt PZHENGST}--{\tt PZHEEVR}--{\tt PZTRSM}.} 

We stress that the slow 
routines {\tt PZHEGST} and {\tt  PZHETRD},
for the reduction to standard and tridiagonal form, respectively, 
were {\em not} used, and instead replaced by the faster  
{\tt PZHENGST} and {\tt PZHENTRD}.
In order to make
use of ScaLAPACK's fast reduction routines,
only the lower 
triangular part of the matrices is referenced and enough memory for a possible
redistribution of the data is provided. 

Elemental (ver.~0.6) -- incorporating PMRRR (ver.~0.6) -- was used for the
EleMRRR timings. In general, since Elemental does not tie 
the algorithmic block size to the distribution block size, 
different block sizes could be used for each of the stages. 
We do not exploit this fact in the reported timings. 
Instead the same block size is used for all stages.  
A block size of around 96 was in all cases optimal, therefore 
experiments were carried out for block sizes of 64, 96 and
128, but only the best timings are reported.\footnote{
The block size for matrix vector products were fixed to 32 in all cases. 
For the biggest matrices in the weak scaling experiment only the block 
size of 32 and 96 were used for ScaLAPACK and EleMRRR,
respectively.}   

Since the timings of the tridiagonal eigenproblem depend on the input data,
so does the 
overall solver. In order to compare fairly different solvers, we fixed
the tridiagonal input matrix by using the 1--2--1 type.  The
performance of every other stage is data independent.
Moreover, since the output of the tridiagonal solvers has to be in a
format suitable for the backtransformation, the MRRR-based routines
have to undergo a data redistribution;  in all the experiments, the timings
for Stage 4 include the cost of the redistribution. 

In the next two subsections we show results for both strong and weak scaling. In
all experiments the number of nodes are increased from 8 to 256. 
Since each node consists of two quad-core 
processors, this corresponds to 64 to 2048 cores.

\subsubsection{Strong Scaling}
\label{sec:strngscal}

\begin{figure}[thb]
   \centering
\includegraphics[scale=.38]{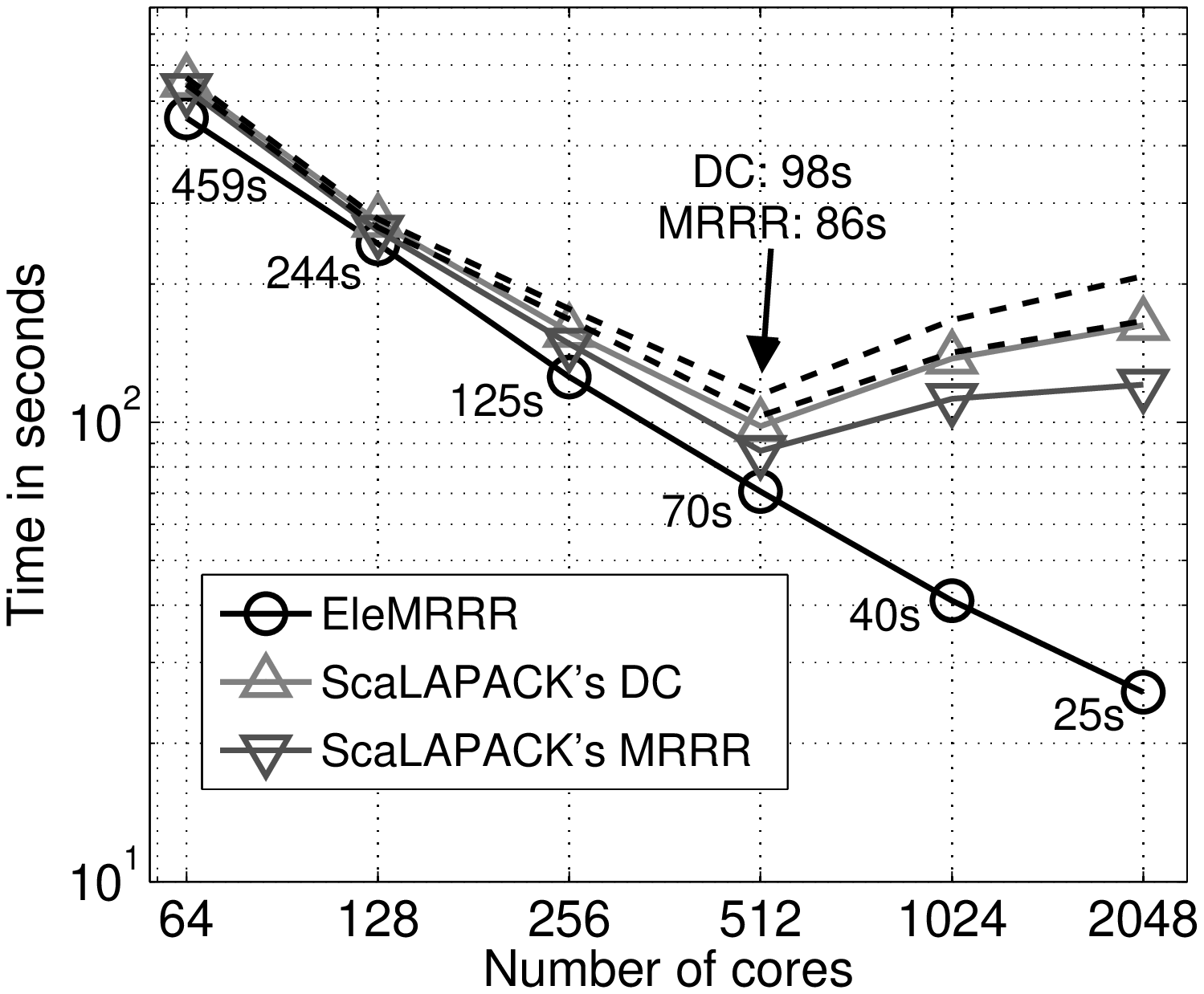}
\includegraphics[scale=.38]{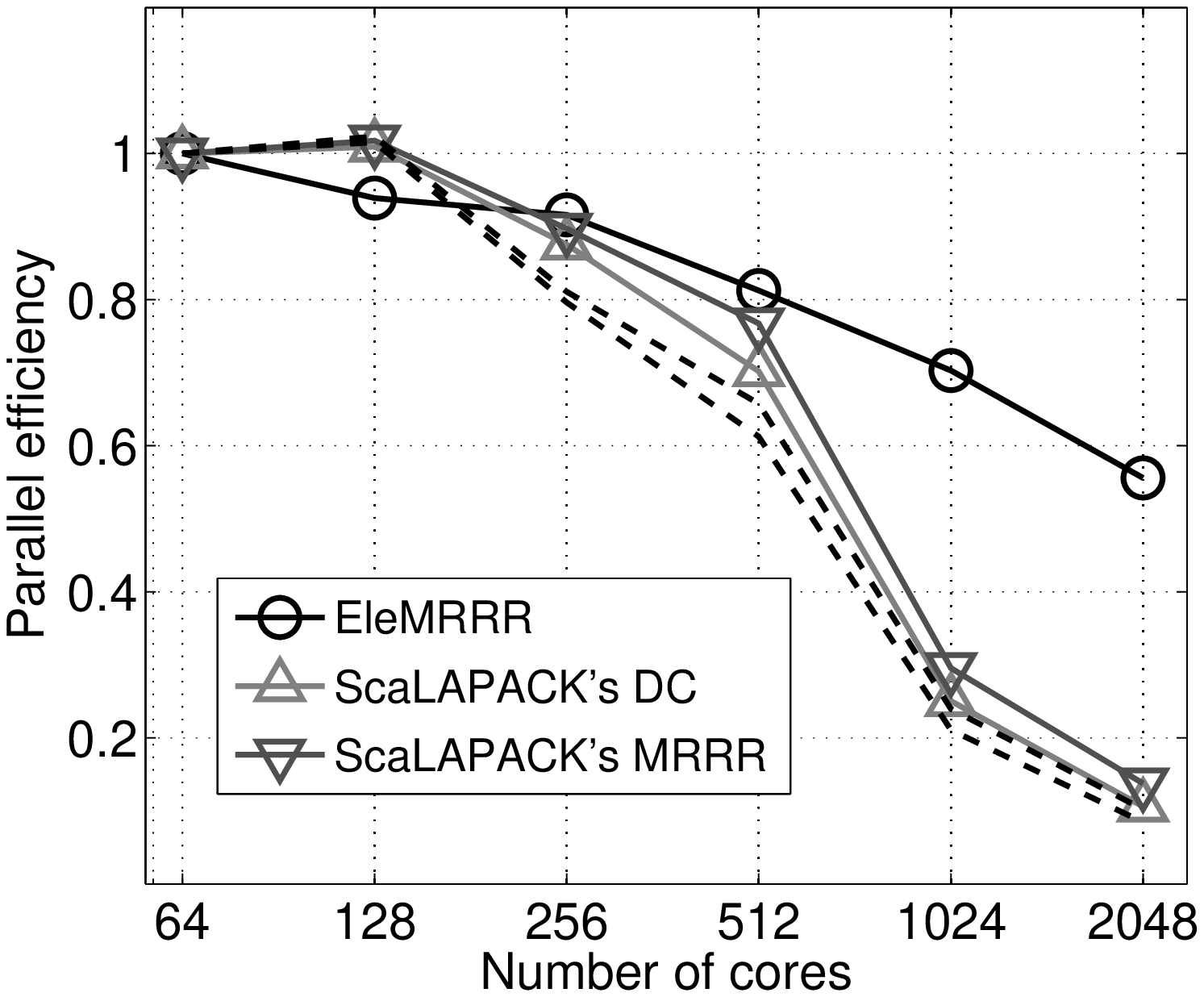}
   \caption{
     Strong scalability for the computation of all eigenpairs.  
     Matrices $A$ and $B$ are of size 20{,}000. 
     The dashed lines refer to ScaLAPACK's solvers when the matrices $A$ and
     $B$ are stored in the upper triangular part; in this scenario, the 
       non-square routines for the reductions are used. 
     {\it
       Left:} Total 
     execution time 
     in a log-log scale. {\it Right:} Parallel efficiency; normalized to the
     execution using 64 cores. 
   }
   \label{fig:time_juropa_strngscal}
\end{figure}

We present timings of EleMRRR for the computation of all the eigenpairs 
of the generalized Hermitian eigenproblem $Ax=\lambda Bx$ for fixed problem
size, $n=20{,}000$. 
The results are displayed in Fig.~\ref{fig:time_juropa_strngscal}. 
As a reference, we have included timings for ScaLAPACK's
solvers.\footnote{We did not investigate the cause for the 
  increased 
  run time of ScaLAPACK using 1024 and 2048 cores. While most subroutines in the
  sequence are slower compared with the run time using 512 cores, {\tt PZHENTRD}
  scales well up to the tested 2048 cores -- see also
  Fig.~\ref{fig:timepzheevd2}.}  
The right side of the figure shows the {\em parallel efficiency} -- defined as $e_p =
(t_{ref} \cdot p_{ref})/(t_p \cdot p)$ -- where $t_p$ denotes the execution
time on $p$ processors, and $t_{ref}$ the execution time on $p_{ref}$
processors. 
The reference time used 8 nodes (64 cores).

Once the proper sequence of routines is rectified, the performance of ScaLAPACK's
solvers is comparable to that of EleMRRR, up to 512
cores (see Fig.~\ref{fig:time_juropa_strngscal}).
For 64 to 512 cores DC is about 10\% to 40\% slower than EleMRRR, while
ScaLAPACK's MRRR is about 7\% to 20\% slower.  The advantage of EleMRRR mainly
comes from the Stages 1, 2 and 6, i.e., those related to the generalized
eigenproblem.  The
timings for the standard problem (Stages 3--5) are nearly identical,
with DC slightly slower than both MRRR-based solutions.

The story for 1024 and 2048 cores changes; the performance of
ScaLAPACK's routines for the generalized eigenproblem drop dramatically.
Compared to DC, 
EleMRRR is about 3.3 and 6.3 times faster; with respect to ScaLAPACK's MRRR
instead, EleMRRR is 2.7 and 4.7 times faster.  The same holds for the
standard eigenproblem, where EleMRRR is about 2.9 and 6.2 times faster than
DC and 1.9 and 3.7 times faster than MRRR.

A study on the Juropa supercomputer suggests that the run time of
applications can be greatly affected by settings of the underlying MPI
implementation~\cite{juropahpc}. In particular, the switch from
static -- the default setting -- to dynamic memory allocation for MPI
connections may result in improved performance.  In regards to the
performance degradation appearing in
Fig.~\ref{fig:time_juropa_strngscal} {\it (left)}, such a switch 
positively impacts some of the stages, including
{\tt PDSTEDC}; on the other hand it gravely worsens other stages, including 
{\tt PDSTEGR}. Overall, ScaLAPACK's DC would only marginally improve,
while  MRRR's performance would greatly deteriorate. 
As a consequence, we employ the default MPI settings in all 
later experiments. 

\begin{figure}[thb]
   \centering
   \includegraphics[scale=.38]{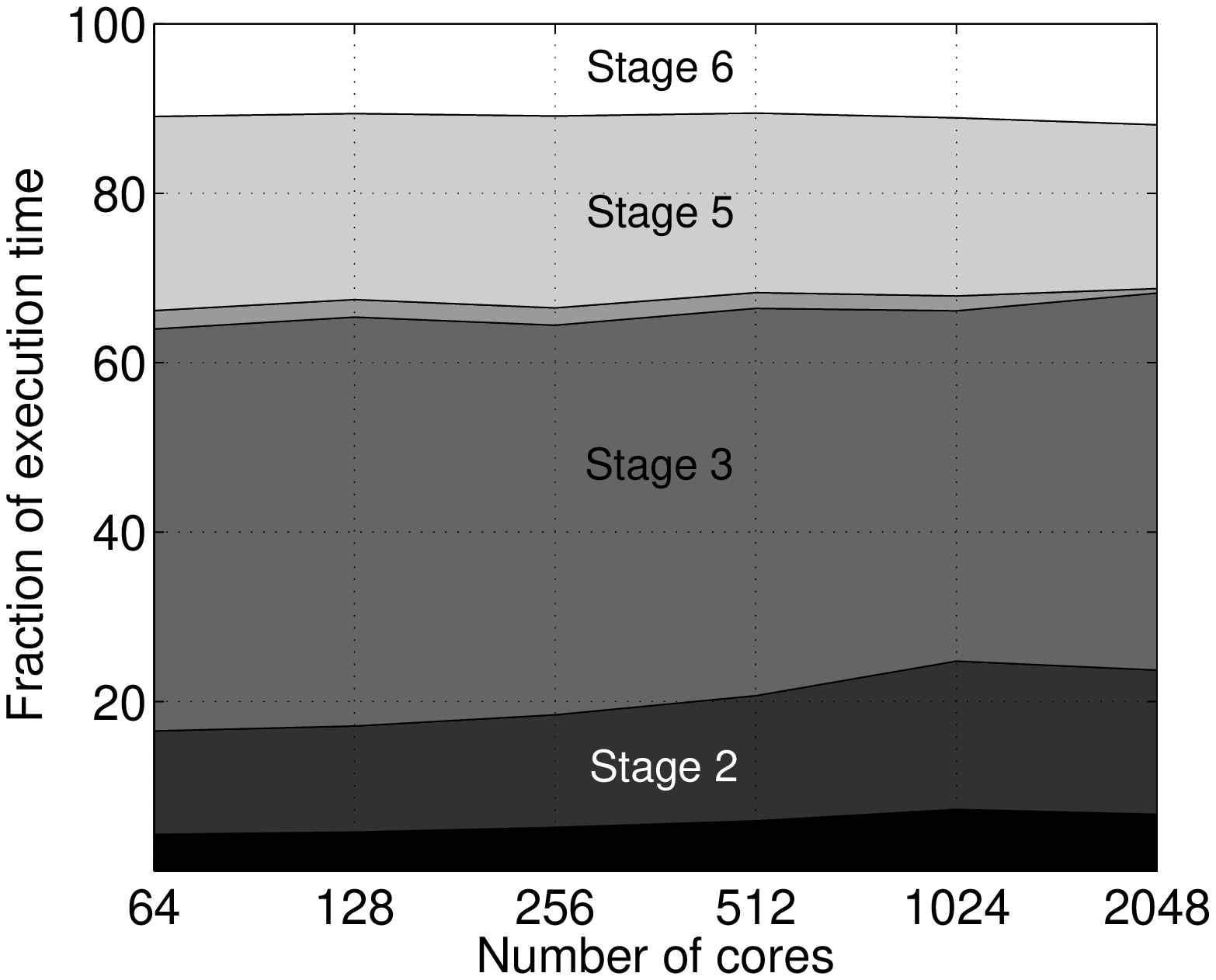} 
\includegraphics[scale=.38]{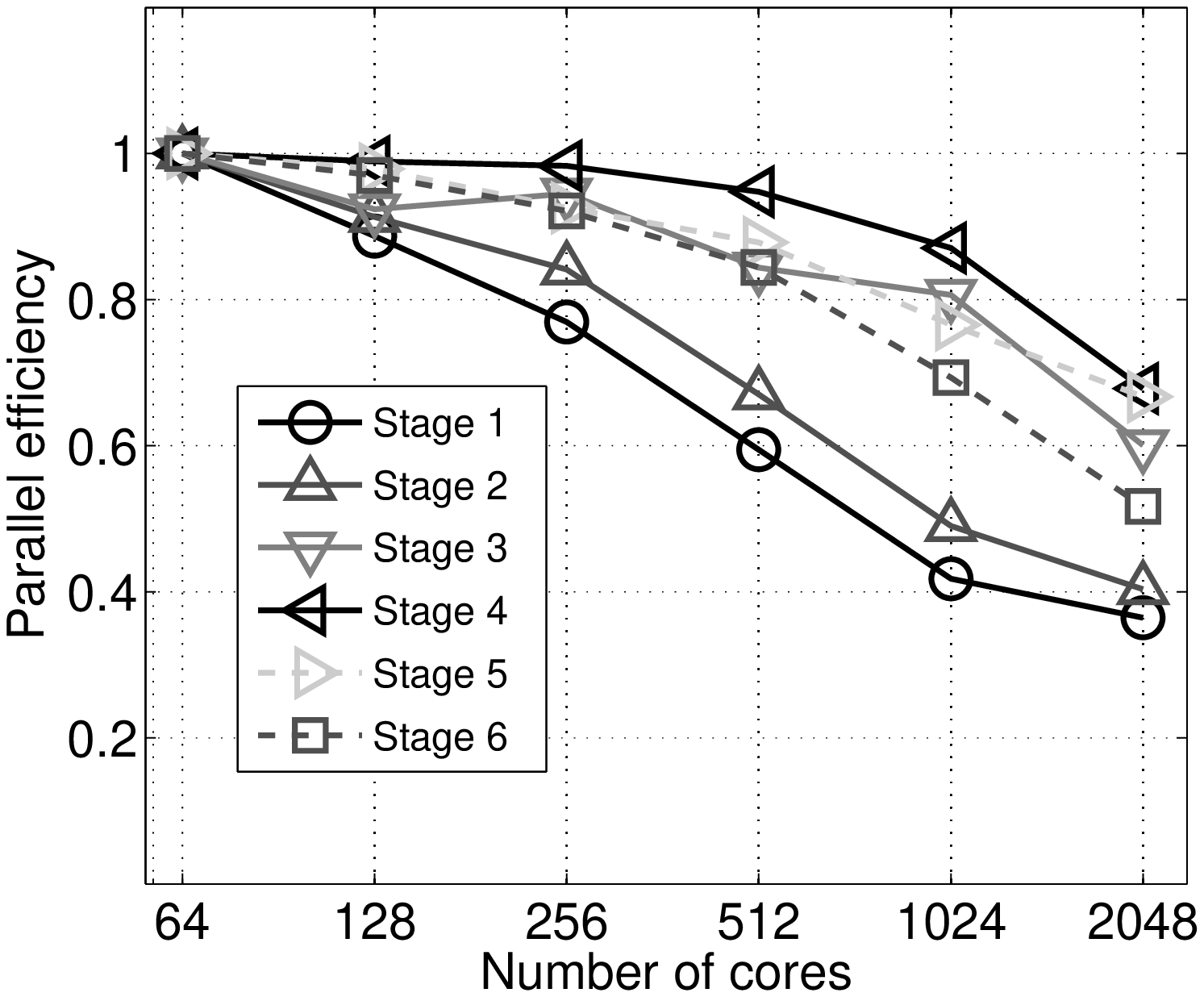}
   \caption{
     EleMRRR's strong scalability for the computation of all
     eigenpairs. Similar results for ScaLAPACK's DC are shown in
     Fig.~\ref{fig:strngscaldcfrac}.   
     {\it Left:} Parallel efficiency
     for all six stages of the computation. Achieving a
     high parallel efficiency is especially important for the stages in which
     most part of the computation is spent. {\it Right:} Fraction of time
     spent in all six stages of the computation. The most time consuming stages
     are the reduction to tridiagonal form, the reduction to standard form,
     and the backtransformation to HEP, while the
     tridiagonal eigensolver is negligible. The time spent in the last three
     stages is proportional to the number of eigenpairs computed.
   }
   \label{fig:frac_juropa_strngscal}
\end{figure}

In Fig.~\ref{fig:frac_juropa_strngscal} we take a closer look at the six
different stages of EleMRRR. 
The left panel tells us that
roughly one third of EleMRRR's execution time -- corresponding to Stages
4, 5 and 
6 -- is proportional to the fraction of computed eigenpairs. Computing a
small fraction of eigenpairs would therefore require roughly about two thirds of
computing the complete decomposition. 
On the right-hand panel of Fig.~\ref{fig:frac_juropa_strngscal}, we report
the parallel efficiency for all six stages separately.  
When analyzed in conjunction with the left figure, this panel indicates if
and when a routine becomes a bottleneck due to bad scaling.

The tridiagonal eigensolver (Stage 4) obtains
the highest parallel efficiency. On the other hand, it 
contributes for less than 2.2\% to the overall run time and is therefore 
negligible in all experiments. 

Up to 1024 cores, ScaLAPACK's MRRR shows a similar behavior: 
the tridiagonal stage
makes up for less than 6\% of the execution time. With 2048 cores instead, the
percentage increases up to 21\%. 
The situation is even more severe for DC, as the fraction spent in the
tridiagonal 
stage increases from about 4.5\% with 64 cores to 41\% with 2048 cores. 
This analysis suggests that the tridiagonal stage,
unless as scalable as the other stages,
will eventually account for a
significant portion of the execution time.




\subsubsection{Weak Scaling}

\begin{figure}[thb]
   \centering
   \includegraphics[scale=.38]{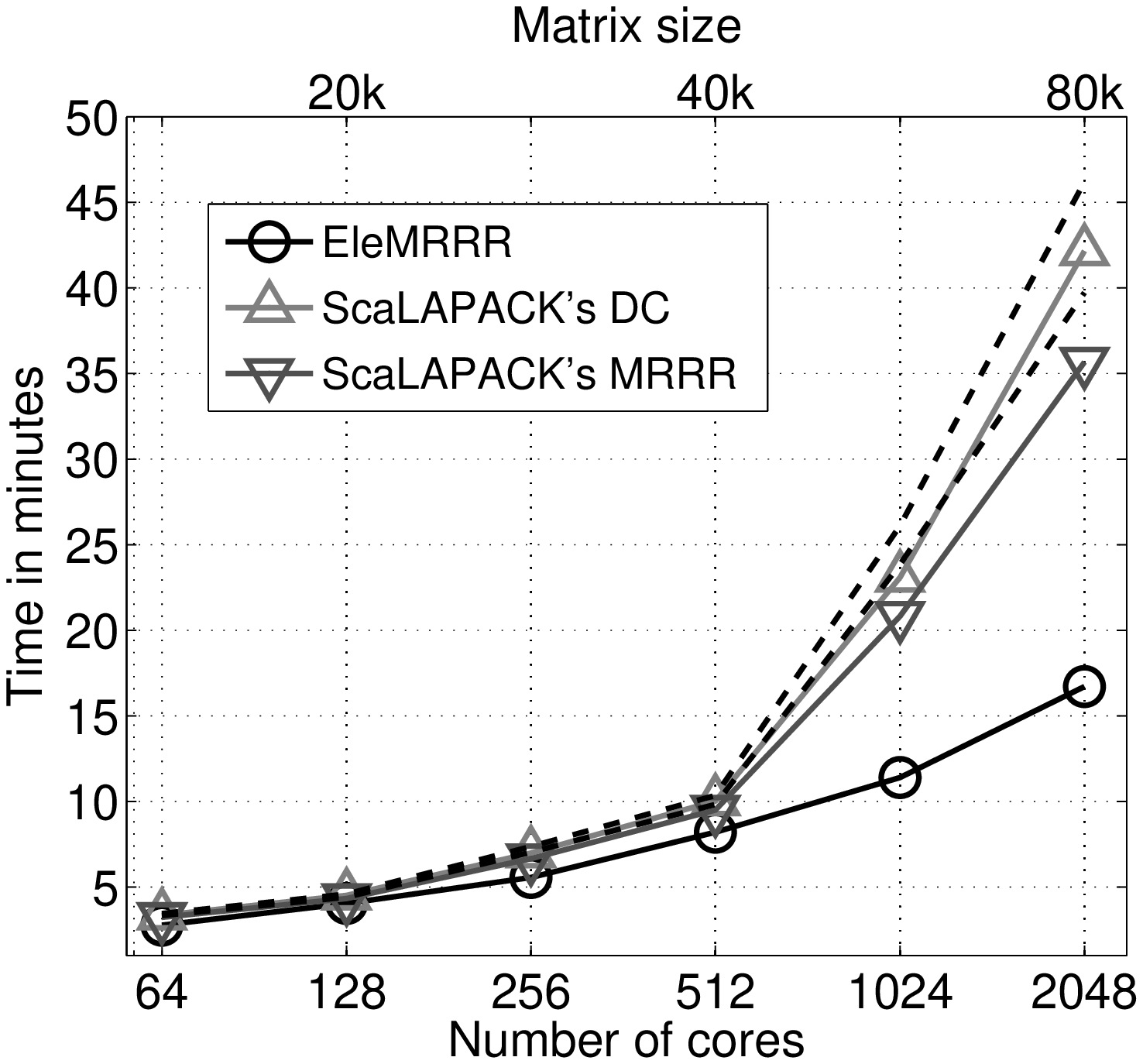} 
\includegraphics[scale=.38]{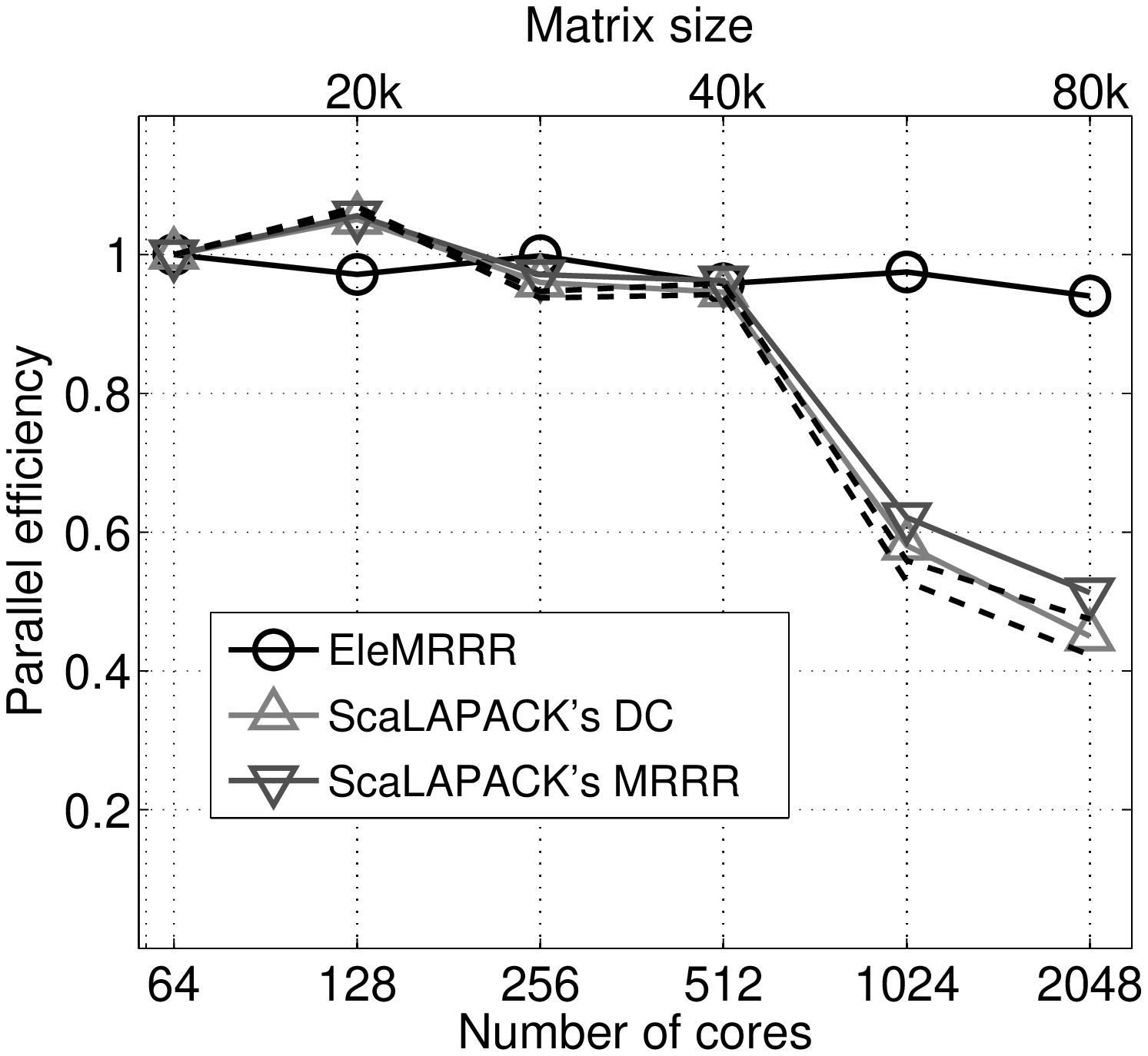} 
   \caption{
     Weak scalability for the computation of all eigenpairs. Matrices $A$
     and $B$ are of varying   
     size. 
     The dashed lines refer to ScaLAPACK's solvers when the matrices $A$ and
     $B$ are stored in the upper triangular part; in this scenario, the 
       non-square routines for the reductions are used. 
     {\it Left:} Total execution 
     time. {\it Right:} Parallel efficiency relative to 64 cores.
     All three routines are fast and efficient
       up to 512 cores. For higher numbers of nodes the
       efficiency of the ScaLAPACK routines drops, while EleMRRR
       retains almost perfect scalability.  }
   \label{fig:time_juropa_weakscal}
\end{figure}

Fig.~\ref{fig:time_juropa_weakscal} illustrates EleMRRR's timings for the
computation of all the eigenpairs of $Ax = \lambda B x$. In this experiment,
the matrix size increases 
(from $14{,}142$ to $80{,}000$)
 together with the number of cores (from 64 to 2048).
The right panel of the figure contains the {\em parallel
efficiency}, defined as $e_p = (t_{ref} \cdot p_{ref} \cdot n^3)/(t_p \cdot p
\cdot n_{ref}^3)$, where all the quantities are like in the previous section and
$n_{ref}$ denotes the smallest matrix size in the experiment ($14{,}142$).

In the tests using 512 cores and less, EleMRRR outperforms ScaLAPACK
only by a small margin, while using 1024 cores and more, the difference
becomes significant. 
The right graph indicates that EleMRRR scales well to large problem sizes
and high number of processes, with parallel efficiency close to one. 
Thanks to its better scalability, for the biggest problem EleMRRR is 2.1
and 2.5 times faster than ScaLAPACK's MRRR and DC, respectively.

\begin{figure}[thb]
   \centering
   \includegraphics[scale=.38]{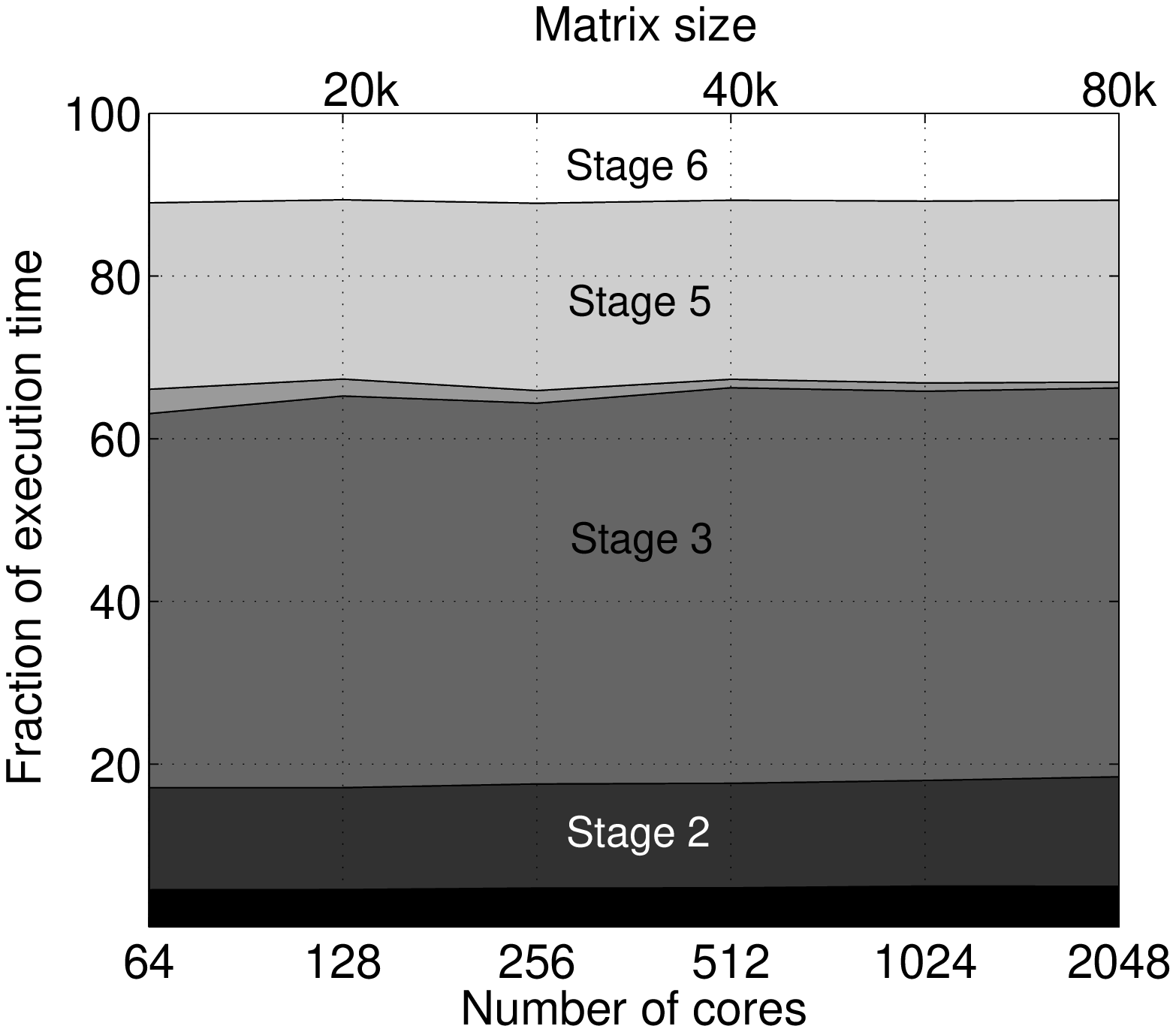} 
\includegraphics[scale=.38]{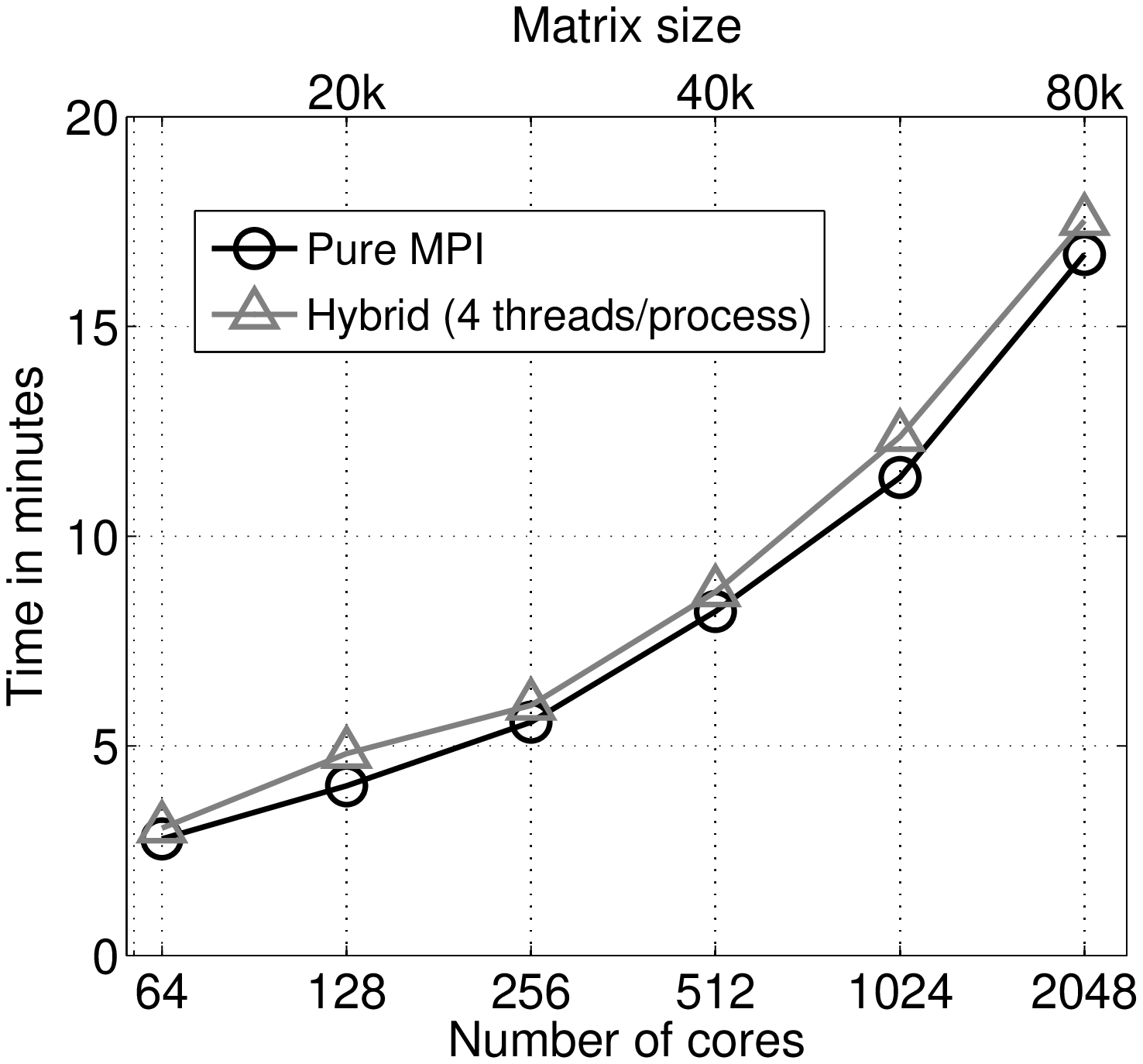} 
   \caption{
     EleMRRR's weak scalability for the computation of all eigenpairs. 
     {\it Left:} Fraction of the execution time spent in the six stages, 
     from bottom to top. 
     {\it Right:} Comparison between a pure MPI execution and a hybrid execution
     using one process per socket with four threads per process.
   }
   \label{fig:frac_juropa_weakscal}
\end{figure}

The execution time is broken down into stages in 
Fig.~\ref{fig:frac_juropa_weakscal} {\it (left)}.
Four comments follow. 
(a) The time spent in PMRRR (Stage 4) is in the range of 
2.5\% to 0.7\% and it is is completely negligible, especially for 
large problem sizes.\footnote{The timings relative to only Stage 4 are
  detailed in Fig.~\ref{fig:timetrdeig}.}
(b) The
timings corresponding to the standard eigenproblem (Stages 3--5)
account for about 72\% of the 
generalized problem's execution time. 
(c) The part of the solver whose execution time is roughly proportional to the
fraction of desired eigenpairs (Stages 4--6)
makes up 32\%--37\% of the 
execution for both the GHEP and HEP. 
(d) No one stage in EleMRRR is becoming a bottleneck, as all of them 
scale equally well.

Fig.~\ref{fig:frac_juropa_weakscal} {\it (right)} shows the
execution of 
EleMRRR using one process per socket with four threads per process. 
The resulting execution time is roughly the same as for the pure MPI execution, 
highlighting the potential of Elemental's hybrid mode. 
Similar experiments with a higher degree of multithreading would positively
affect PMRRR, but not the reduction to tridiagonal form.

\subsection{Jugene}

In this section, we verify the prior results obtained on Juropa for
a different architecture -- namely, the {\it BlueGene/P} installation {\it
  Jugene}. 
Jugene consists of 73{,}728 nodes, each of which is equipped with 2 GB of
memory and a quad core PowerPC 450 processor running at 850 MHz.

All routines were compiled using the {\it IBM XL} compilers (ver.~9.0) in
combination with the vendor tuned 
{\it IBM} MPI library. 
We used a square processor grid $P_r = P_c$ whenever possible
and $P_r = 2 P_c$ otherwise.\footnote{As noted in
    Section~\ref{juropa}, $P_c \approx P_r$ or the 
  largest square grid possible should be
  preferred. These choices do
  not affect the qualitative behavior of our performance results.} Similarly, ScaLAPACK (ver.~1.8) in 
conjunction 
with the vendor-tuned BLAS included in the ESSL library (ver.~4.3) was used
throughout.\footnote{At the time of writing, ScaLAPACK's
  up-to-date version was 1.8. The current version, 2.0, presents no
  significant changes in the tested routines.} In contrast to the Juropa
experiments, we 
concentrate on the weak scalability of the {\it symmetric-definite} generalized
eigenproblem. Therefore ScaLAPACK's DC timings correspond to the 
sequence of routines {\tt PDPOTRF}--{\tt PDSYNGST}--{\tt PDSYNTRD}--{\tt PDSTEDC}--{\tt
  PDORMTR}--{\tt PDTRSM}. 
Accordingly, ScaLAPACK's MRRR corresponds to the same sequence of routines
with {\tt PDSTEDC} replaced by {\tt PDSTEGR}.\footnote{As {\tt
    PDSTEGR} is not contained in ScaLAPACK, it corresponds to the sequence
  {\tt PDPOTRF}--{\tt PDSYNGST}--{\tt PDSYEVR}--{\tt PDTRSM}.} 
In both cases, a block size of
48 was found to be nearly optimal 
and used in all experiments. As already 
explained in Section~\ref{juropa}, we avoided the use of the routines {\tt
  PDSYGST} and {\tt PDSYTRD} for the reduction to standard and tridiagonal
form, respectively.

For EleMRRR's timings we used Elemental (ver.~0.66), which integrates
PMRRR (ver.~0.6). A block size of 96 was identified as nearly optimal and
used for all experiments.\footnote{
The block size for matrix vector products, which does not have a significant
influence on the performance, was fixed to 64 in all cases.}


\subsubsection{Weak Scaling}
\begin{figure}[ht]
	\begin{center}
		\includegraphics[scale=.38]{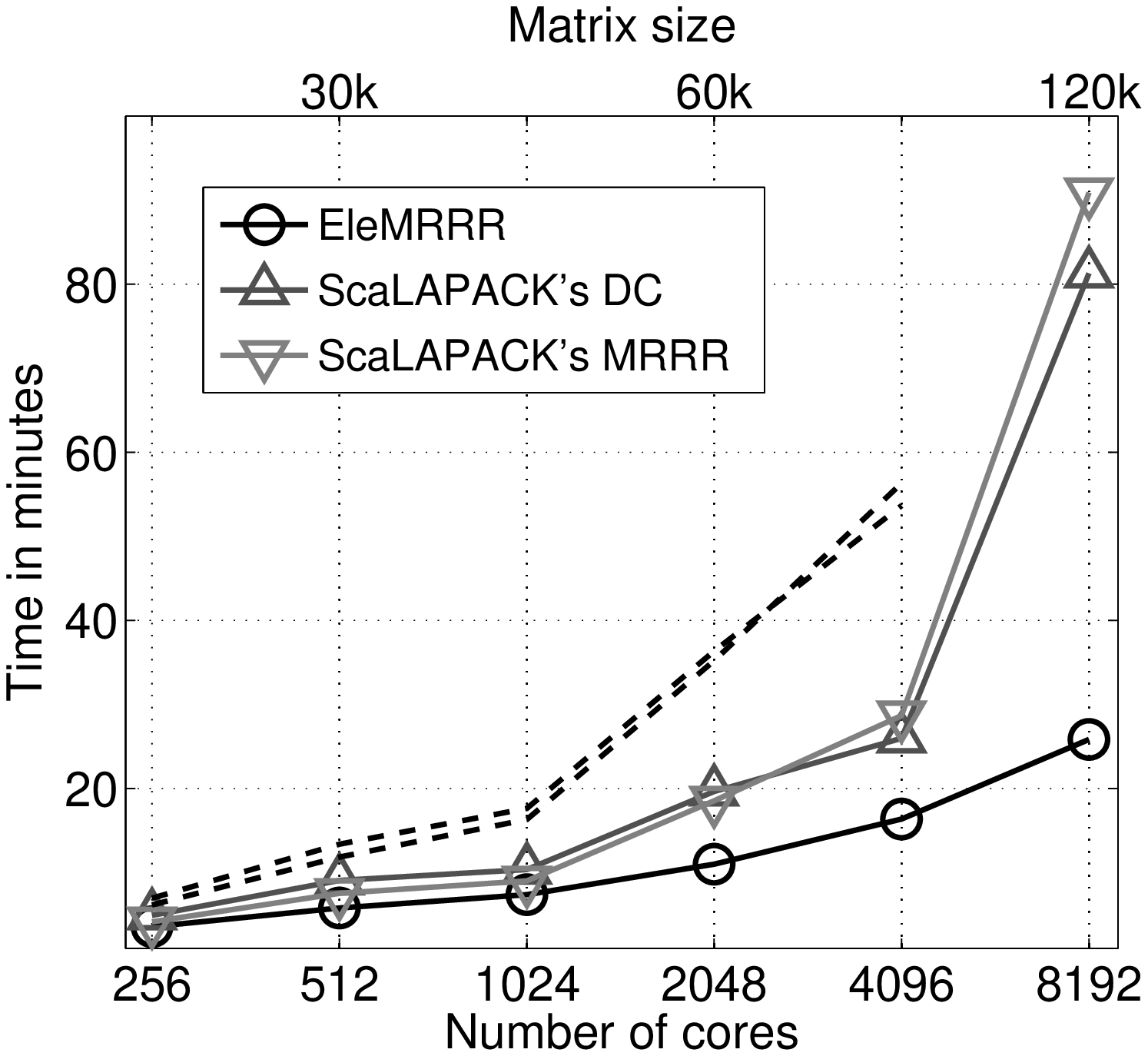}
		\includegraphics[scale=.38]{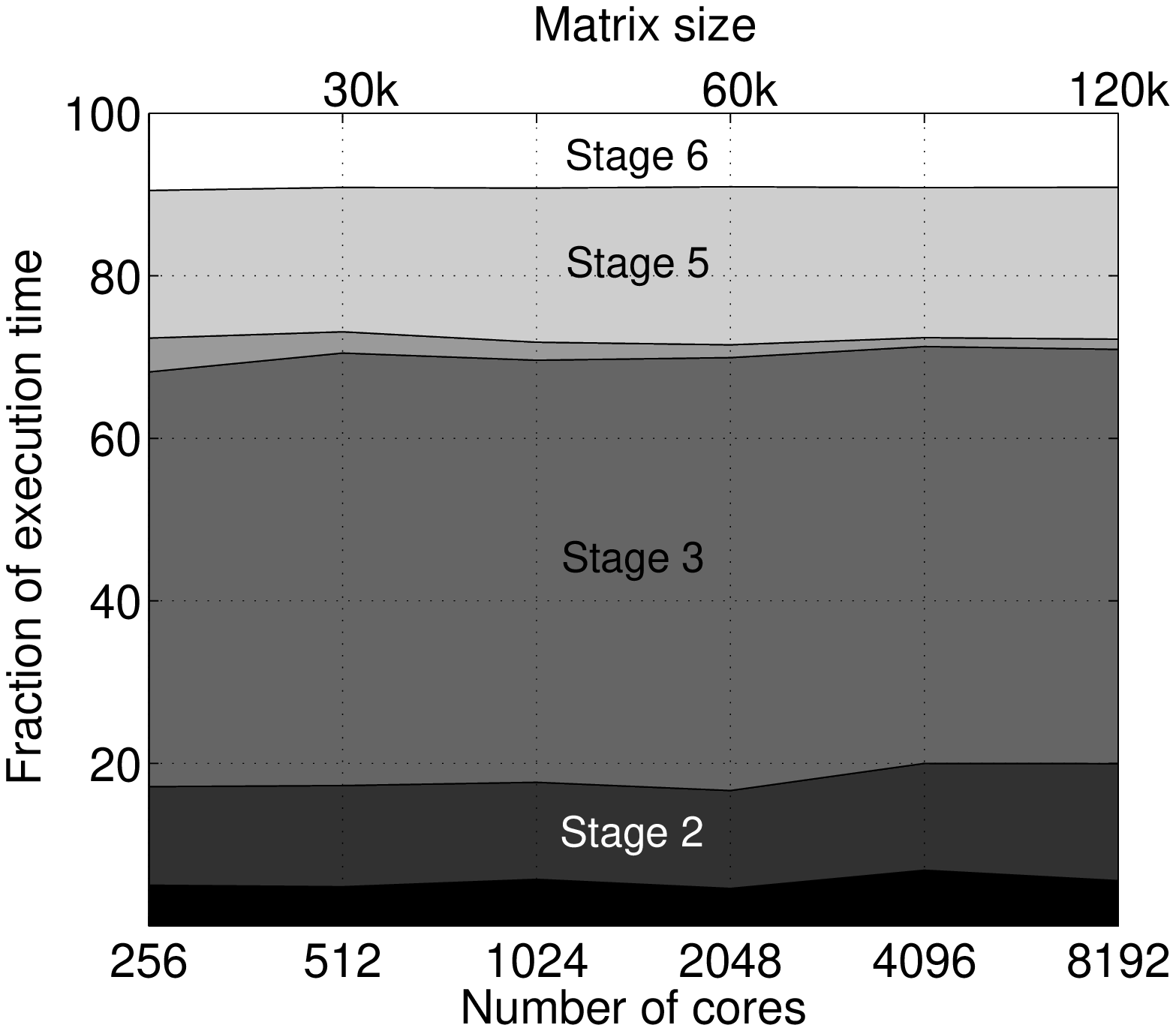}
	\end{center}
	\caption{
          Weak scalability for computing all eigenpairs of $Ax = \lambda B
          x$. 
     The dashed lines refer to ScaLAPACK's solvers when the matrices $A$ and
     $B$ are stored in the upper triangular part; in this scenario, the 
       non-square routines for the reductions are used. 
   {\it Left:} Total execution time. EleMRRR is the fastest solver
          across the board. {\it Right:} Fraction of the
          execution time spent in the six stages, from bottom to top.} 
	\label{fig:bgp:weak}
\end{figure}

In the left panel of Fig.~\ref{fig:bgp:weak} we present EleMRRR's timings
for the computation of all 
eigenpairs of the generalized problem in the form of $Ax = \lambda Bx$. While
the size of the test matrices ranges from 
$21{,}214$ to $120{,}000$, the number of nodes increases from $64$ to $2{,}048$
($256$ to $8{,}192$ cores). In the right panel, the execution time is broken down
into the six stages of the generalized eigenproblem.

Both graphs show a similar behavior to the experiments performed on
Juropa. In all experiments EleMRRR outperforms both the
ScaLAPACK's solvers. 
Most importantly, ScaLAPACK again suffers from a breakdown in scalability. 

When analyzing the results of Fig.~\ref{fig:bgp:weak} for
ScaLAPACK's solver we the following observations. (a) 
The scalability issues can be attributed mainly to
Stages 1, 2 and 4. In particular, for the largest experiment
ScaLAPACK's reduction to standard form (28 minutes) and MRRR
(25 minutes) each exceed the time that EleMRRR spends for the entire
problem. 
(b) 
Although ScaLAPACK's tridiagonal eigensolver is usually not very time
consuming, for highly parallel systems it might become the bottleneck.
As an example, in our experiment on 8,192 cores the tridiagonal
problem accounts for 54\% (MRRR) and 33\% (DC) of the total execution time
of the standard eigenproblem.  
(c) Due to better
  scalability, DC becomes faster than ScaLAPACK's MRRR.

We conclude with four comments regarding EleMRRR’s behavior. 
 (a)
While stage 4 of ScaLAPACK's MRRR and DC take up to 
28\% and 20\% of the  total
execution time, respectively, PMRRR accounts for less than 4\%. In
particular, for the largest problem 25 minutes were spent in
ScaLAPACK's MRRR, whereas PMRRR required only 20 seconds. 
In all experiments
PMRRR's execution time was negligible. 
(b) The
timings corresponding to the standard eigenproblem account for 70\%--74\%  
 of the generalized problem's execution time.
(c) The part which is roughly proportional to the
fraction of desired eigenpairs 
makes up 26\%--32\% of both the generalized and
standard problem. (d) All the six stages
scale equally well and no computational bottleneck emerges.

\section{Conclusions}
\label{conclusions}


Our study of dense large-scale generalized and standard eigenproblems
was motivated by performance problems of commonly used routines in the
ScaLAPACK library. In this paper we identify such problems and provide
clear guidelines on how to circumvent them: by invoking suitable routines
with the right settings, the users can assemble solvers faster than
those included in ScaLAPACK.

The main contribution of the paper lies in the introduction of Elemental's
dense eigensolvers and our tridiagonal eigensolver, PMRRR. Together, they
provide a set of routines -- labeled 
EleMRRR -- for large-scale eigenproblems. These
solvers are 
part of the publicly available Elemental library and make use of
PMRRR, a parallel version of the MRRR 
algorithm for computing all or a subset of eigenpairs of tridiagonal
matrices. PMRRR supports pure message-passing, pure multithreading,
as well as hybrid executions, and our experiments indicate that it is among 
the fastest and most scalable tridiagonal eigensolvers currently available.


In a thorough performance study on two state-of-the-art
supercomputers, we compared EleMRRR with the solvers built within the ScaLAPACK
framework according to our guidelines.
For a modest amount of parallelism, ScaLAPACK's solvers obtain
results comparable to EleMRRR, provided the fastest routines with suitable
settings are invoked. In general, EleMRRR attains the best
performance and obtains the best scalability of all solvers.



\section*{Acknowledgments}
This article could have not materialized without the work and help of Jack
Poulson (UT Austin, Texas, USA). 
We are also grateful for the comments of
Robert van de Geijn (UT Austin, USA), Edoardo Di Napoli (J\"ulich
Supercomputing Center, Germany), and the anonymous reviewers. 
Furthermore, the authors would
like to thank Inge Gutheil and Stefan Bl\"ugel (Research Center 
J\"ulich, Germany).
Finally, sincere thanks are extended to the J\"ulich Supercomputing Center
for granting access to Juropa and Jugene, allowing us to complete all the
experiments.

\bibliographystyle{siam}
\bibliography{elemrrr}


\end{document}